\documentclass[aps,prl,twocolumn,superscriptaddress,longbibliography]{revtex4-1}
\usepackage{graphicx}
\usepackage{amsmath}
\usepackage{amssymb}

\usepackage{amsmath}
\usepackage{pifont}
\usepackage{amssymb}
\usepackage{latexsym}
\usepackage{dsfont }
\usepackage{graphicx}
\usepackage{float}
\usepackage{color}
\usepackage{multirow}
\usepackage{verbatim}

\usepackage{pdfpages}

\makeatletter
\AtBeginDocument{\let\LS@rot\@undefined}
\makeatother

\renewcommand{\arraystretch}{1.5}

\def\bk{{\bold{k}}}

\def\up{{\uparrow}}
\def\dn{{\downarrow}}

\def\m1{{^{-1}}}

\begin{document}


\title{The effects of strain in multi-orbital superconductors: 
the case of Sr$_2$RuO$_4$}

\author{Sophie Beck}
\affiliation{Center for Computational Quantum Physics, Flatiron Institute, 162 5th Avenue, New York, NY 10010, USA}

\author{Alexander Hampel}
\affiliation{Center for Computational Quantum Physics, Flatiron Institute, 162 5th Avenue, New York, NY 10010, USA}

\author{Manuel Zingl}
\affiliation{Center for Computational Quantum Physics, Flatiron Institute, 162 5th Avenue, New York, NY 10010, USA}

\author{Carsten Timm}
\affiliation{Institute of Theoretical Physics, Technische Universität Dresden, 01062 Dresden, Germany}
\affiliation{Würzburg-Dresden Cluster of Excellence ct.qmat, Technische Universität Dresden, 01062 Dresden, Germany}

\author{Aline Ramires}
\affiliation{Paul Scherrer Institut, CH-5232 Villigen PSI, Switzerland}

\begin{abstract}
Uniaxial strain experiments have become a powerful tool to unveil the character of unconventional phases of electronic matter. Here we propose a combination of the superconducting fitness analysis and density functional theory (DFT) calculations in order to dissect the effects of strain in complex multi-orbital quantum materials from a microscopic perspective. We apply this framework to the superconducting state of Sr$_2$RuO$_4$, and argue that the recently proposed orbitally anti-symmetric spin-triplet (OAST) order parameter candidate has unique signatures under strain which are in agreement with recent observations. In particular, we can account for the asymmetric splitting of the critical temperatures for compressive strain along the $\langle 100\rangle$ direction, and the reduction of the critical temperatures for compressive strain along the $\langle 001\rangle$ and $\langle 110\rangle$ directions with a single free parameter.
\end{abstract}

\date{\today}

\maketitle

The recent development of strain devices suitable for a variety of experimental probes has opened a new direction of investigation of complex quantum materials \cite{Hicks:2014}. The application of uniaxial strain along different directions allows for the selective reduction of spatial symmetries, which is key to uncover the character of the underlying phases of matter in a variety of magnetic and superconducting materials \cite{Brodskye:2017,Park2:2018, Bohmer:2017, Kim:2018, Sun:2019}. Strain can tune material parameters such as orbital occupation \cite{Pfau:2019}, Fermi surface geometry \cite{Hsu:2016}, and topological properties \cite{Mutch:2019}. Given the complexity of most of the functional materials available today, a clear understanding of the effects of strain from a microscopic perspective is highly desirable.

Here we take as an example Sr$_2$RuO$_4$, a material whose superconducting order parameter has been the focus of debate for more than 25 years \cite{Mackenzie:2017}. This system has recently been investigated by thermodynamic \cite{Watson:2018,Steppke:2017}, transport \cite{Barber:2018}, angle-resolved photoemission spectroscopy (ARPES) \cite{Sunko:2019}, nuclear magnetic resonance (NMR) \cite{Pustogow:2019}, and muon spin relaxation ($\mu$SR) \cite{Grinenko:2021} experiments under strain, giving us important hints on the nature of  the superconducting state in this material.

The best contenders for the superconducting state of Sr$_2$RuO$_4$ are chiral order parameters \cite{Rice:1995, Kallin:2009,Kallin:2012,Kallin:2016}. Chiral superconductivity is supported by several experimental probes: polar Kerr rotation experiments reveal time-reversal symmetry breaking (TRSB) at the superconducting critical temperature $T_c$ \cite{Xia:2006}, ultrasound attenuation experiments indicate a two-component order parameter \cite{Lupien:2001,Ghosh:2020}, and the study of junctions suggest the presence of superconducting domains \cite{Kidwingira:2006,Nakamura:2013}. Also, recent $\mu$SR measurements under compressive strain along the $\langle 100 \rangle$ direction show a clear splitting between the superconducting critical temperature $T_c$ and the temperature below which TRSB  is observed ($T_{\text{TRSB}}$) \cite{Grinenko:2021}. A splitting is expected for a chiral superconductor, but some of the features observed experimentally cannot be accounted for by a simple phenomenological Ginzburg-Landau theory. First, the transition seems to be rather asymmetric concerning the evolution of the two temperatures: $T_c$ is strongly enhanced, while $T_{\text{TRSB}}$ remains almost unchanged up to uniaxial strains of about $1\%$ \cite{Grinenko:2021}. Second, the evolution of $T_c$ is strongly non-linear, and there is no observable cusp of $T_c$ around zero strain \cite{Watson:2018}. In addition, recent experiments indicate that compressive strain along the $\langle 001 \rangle$ direction causes a reduction of $T_c$ \cite{Jerzembeck:2021}. This behavior is in contradiction with the expected enhancement based on a weak-coupling scenario with an increased density of states (DOS) given the proximity to the van Hove singularity. Furthermore, experiments under compressive strain along the $\langle 110 \rangle$ direction indicate a mild suppression of the critical temperature \cite{Hicks:2014110}.

The original proposal of a chiral superconducting state for Sr$_2$RuO$_4$ suggested a $p$-wave triplet state with \textbf{d}-vector along the $z$-direction \cite{Rice:1995}. This state is hard to reconcile with new NMR experiments \cite{Pustogow:2019}: the reduction of the Knight shift for in-plane fields is expected only for singlet or triplet states with an in-plane d-vector. Furthermore, the standard chiral $p$-wave state has difficulties accounting for thermodynamic \cite{Kittaka:2018} and transport \cite{Hassinger:2017} experiments indicating the presence of gap nodes. In this context, some of us have recently proposed a new order parameter candidate: an even-parity pseudospin-singlet chiral superconducting state with horizontal line nodes \cite{Suh:2020}. In the microscopic orbital basis, this superconducting state in the $E_g$ channel is an $s$-wave orbital-antisymmetric spin-triplet (OAST) which can be stabilized  by local interactions in the presence of a large effective Hund coupling and realistic three-dimensional spin-orbit coupling \cite{Suh:2020}. 

Here we investigate the effects of strain on local order parameters with $E_g$ symmetry. We conclude that the concept of \emph{superconducting fitness} \cite{Ramires:2016, Ramires:2017, Ramires:2018,Ramires:2021} in conjunction with density functional theory (DFT) calculations can clarify the origin of the unusual features of the evolution of $T_c$ and $T_{\text{TRSB}}$ under strain along different directions in a consistent manner. For that, we use the evolution of the normal state band structure from DFT and a single free parameter. Our analysis stems from a microscopic perspective and captures qualitatively new effects beyond a naive Ginzburg-Landau construction. In particular, we discuss the microscopic origin of the asymmetry of $T_c$ and $T_{\text{TRSB}}$ under strain along the $\langle 100 \rangle$ direction, and how strain along the $\langle 001 \rangle$ direction reduces $T_c$, even though the DOS is enhanced. Moreover, we discuss the behavior of the transition temperatures under strain along the $\langle 110 \rangle$ direction.

The simplest form of superconducting fitness measures was introduced in the context of two-orbital models \cite{Ramires:2018}, therefore here we focus on reduced models for Sr$_2$RuO$_4$ along the $k_xk_z$ and $k_yk_z$ planes, both with $D_{2h}$ symmetry (from now on we refer to these planes as $XZ$ and $YZ$, respectively). Note that there are no two-dimensional irreducible representations (irreps) in $D_{2h}$, so at first sight, it seems that we have lost the discussion about the splitting of the degenerate superconducting transitions in $E_g$ for the complete model with $D_{4h}$ symmetry. Here, the reduced models along the  $XZ$ and  $YZ$ planes should be seen as degenerate. Their inequivalence under strain is a manifestation of the symmetry breaking.

Sufficiently far from the Brillouin-zone diagonals ($k_y=\pm k_x$), the bands close to the Fermi energy are dominated by only two of the Ru $t_{2g}$-orbitals. For concreteness, here we consider the $XZ$ plane, dominated by the $d_{xz}$ and $d_{xy}$ orbitals. Projecting into this subspace, we obtain the following effective two-orbital Hamiltonian
\begin{eqnarray}
\mathcal{H}_0^{XZ}=\sum_{\bk}\Psi^\dagger_\bk \hat{H}_0^{XZ}(\bk) \Psi_\bk,
\end{eqnarray}
in the basis $\Psi^\dagger_\bk = (c_{\bk,xz\up}^\dagger,\allowbreak c_{\bk,xz\dn}^\dagger,\allowbreak
c_{\bk,xy\up}^\dagger,\allowbreak c_{\bk,xy\dn}^\dagger)$, with
\begin{eqnarray}\label{Eq:H02orb}
\hat{H}_{0}^{XZ}(\bk) = \sum_{a,b =0}^3 \tilde{h}_{ab}^{XZ}(\bk)\, \hat{\tau}_a \otimes \hat{\sigma}_b,
\end{eqnarray}
where  the $\tilde{h}_{ab}^{p}(\bk)$ are real even functions of momentum labelled by indexes $(a,b)_p$,  with $a$ and $b$ corresponding to $\hat{\tau}_a$ and $\hat{\sigma}_b$, Pauli matrices encoding the orbital and the spin degrees of freedom, respectively ($\hat{\sigma}_0$ and $\hat{\tau}_0$ are identity matrices), and $p=\{XZ,YZ\}$ corresponds to the plane.
There are, in principle, 16 functions labelled as $(a,b)_{XZ}$, but in presence of time-reversal and inversion symmetries these are constrained to only six: $(0,0)_{XZ}$, $(3,0)_{XZ}$, and $(2,1)_{XZ}$ in $A_{g}$, the first two associated with intra-orbital hopping and the last with atomic spin-orbit coupling (SOC); $(2,2)_{XZ}$ in $B_{1g}$ and $(2,3)_{XZ}$ in $B_{2g}$, both associated with momentum-dependent SOC; and $(1,0)_{XZ}$ in $B_{3g}$ associated with inter-orbital hopping. Analogously, we parametrize the local gap matrices in the orbital basis as 
\begin{eqnarray}
\hat{\Delta}^{XZ}= d_0\, \hat{\tau}_a \otimes \hat{\sigma}_b\, (i\hat{\sigma}_2),
\end{eqnarray}
where $d_0$ is the order parameter amplitude.
Similar construction holds for the $YZ$ plane. The derivation of the terms in the normal state Hamiltonian, labelled as $(a,b)_{p}$, and order parameters, labelled as $[a,b]_{p}$, are given in detail in the Supplemental Material (SM) \cite{SM}, which also includes references \cite{Ramires:2019,Vogt:1995,Barber:2019,Blaha:2019,Perdew:1996,Marzar:1997,Souza:2001,Kunes:2010,Pizzi:2020,Parcollet:2015,Giannozzi:2009,Garrity:2014,Min:2006,Sigrist:1991}.

\begin{center}
\begin{figure}[t]
\includegraphics[width=\linewidth, keepaspectratio]{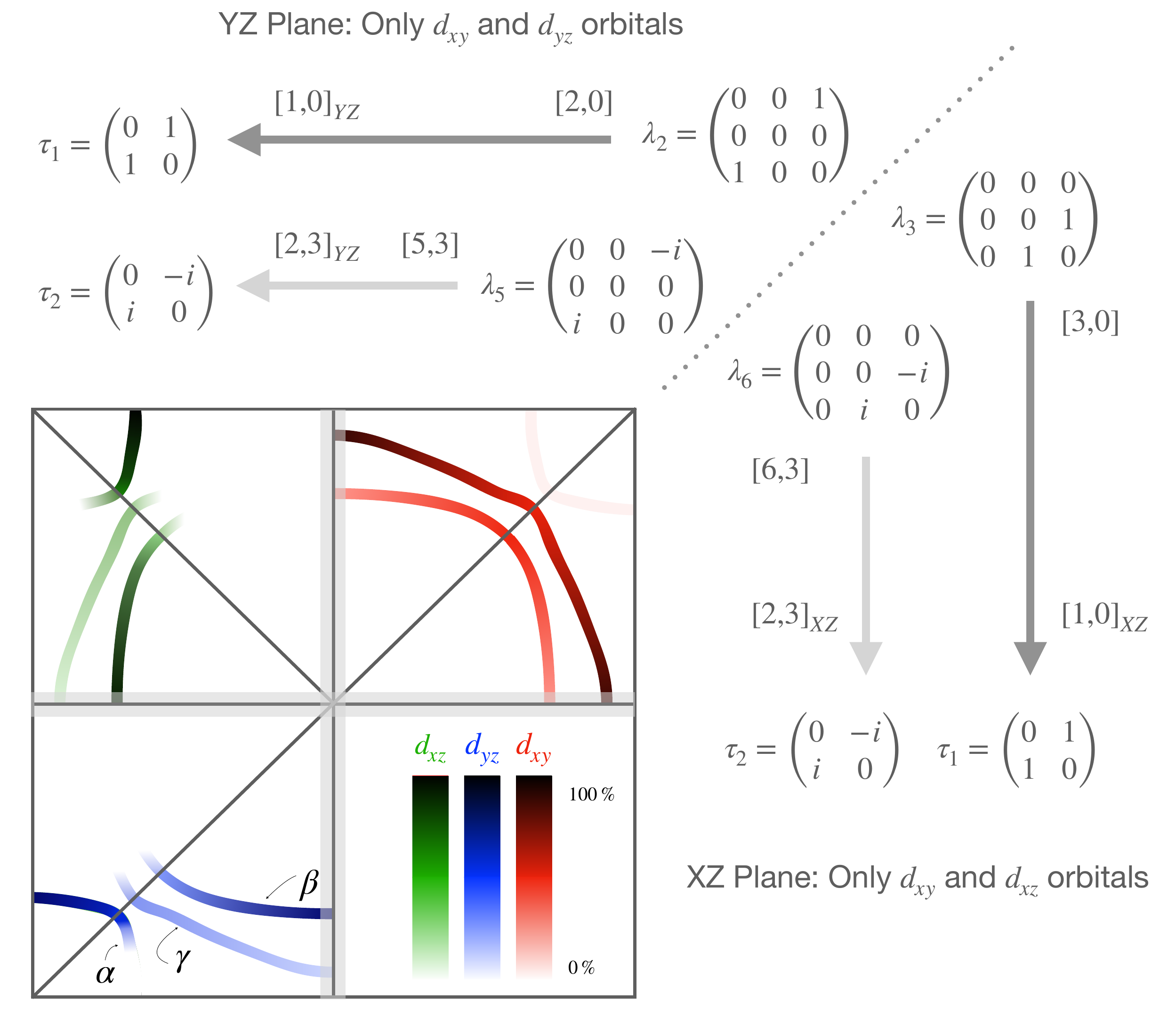}
\caption{The central panel schematically indicates the orbital content over the Fermi surfaces of Sr$_2$RuO$_4$ in the $k_z$ = 0 plane. Highlighted in gray are the XZ and YZ planes discussed in this work. The color red (blue, green) encode the orbital $d_{xy}$ ($d_{yz}$, $d_{xz}$) content as low (bright) or high (dark) color in each Fermi surface sheet (labelled as $\alpha$, $\beta$, and $\gamma$). The $\lambda$-matrices make explicit the orbital configuration of the pairs for the $\{[5,3],[6,3]\}$ and $\{[3,0],-[2,0]\}$ order parameters with $E_g$ symmetry. These are written in the orbital basis $(d_{yz},d_{xz},d_{xy})$.  The long arrows indicate the mapping from the superconducting order parameter components $[a,b]$ in $E_g$ for the three-dimensional model with $D_{4h}$ symmetry to $[a,b]_p$ for the two-dimensional models along the $XZ$ and $YZ$ planes.}
\label{Fig:Corr}
\end{figure}
\end{center}

From here on, we focus on even parity two-component order parameters with $E_g$ symmetry, connecting to the OAST order parameter proposed in \cite{Suh:2020}, which reconciles several experimental observations. In the complete description of Sr$_2$RuO$_4$ as a three-dimensional three-orbital model (see details in the SM \cite{SM}), the order parameter is dominated by the basis matrices $\{[5,3],[6,3]\}$. Once we project the complete model into the $XZ$ or $YZ$ planes, we identify $[5,3] \rightarrow [2,3]_{YZ}$ and $[6,3] \rightarrow [2,3]_{XZ}$. A subdominant contribution to this symmetry channel comes from $\{[3,0],-[2,0]\}$, which in the projected models are identified as  $[2,0] \rightarrow [1,0]_{YZ}$ and $[3,0] \rightarrow [1,0]_{XZ}$. These are the only $E_g$ order parameters we can discuss within the projected models. The quantitative analysis below is done within the two-orbital models, and we use the correspondences summarized in Fig. \ref{Fig:Corr} to connect the results to the original 3-orbital problem with $D_{4h}$ symmetry. 

Within the standard weak-coupling assumptions, the superconducting critical temperature for a two-orbital system under strain can be written as \cite{Ramires:2018}
\begin{eqnarray}\label{Eq:Tc}
T_c (s) &\approx&  \frac{4 e^\gamma}{\pi}\frac{\omega_C}{2 } \exp{\left[\left(-\frac{1}{2|v|}-\delta(s) \right)\frac{1}{\alpha(s)}\right]},
\end{eqnarray}
where $\gamma$ is the Euler constant, $\omega_C$ is a characteristic energy cutoff, $|v|$ is the magnitude of the attractive interaction in the symmetry channel of interest (the last two assumed to be strain independent). The superconducting fitness functions are written as
\begin{eqnarray}
\alpha(s) = \frac{1}{16}\sum_a N_a(0,s) \langle ||\hat{F}_A(\bk_{Fa},s)||^2\rangle_{\text{FS}_a},
\end{eqnarray}
and
\begin{eqnarray}
\delta(s) = \frac{\omega_C^2}{32} \sum_a  N_a(0,s)  \left\langle \frac{||\hat{F}_C(\bk_{Fa},s)||^2}{q(\bk_{Fa})^2} \right\rangle_{\!\!\text{FS}_a},
\end{eqnarray}
where $q(\bk)= \epsilon_a(\bk)-\epsilon_b(\bk)$ corresponds to the energy difference between the two bands. The sum over $a$ corresponds to the sum over the Fermi surfaces with DOS $N_a(0,s)$ at the Fermi level for strain $s$, and $\langle ... \rangle_{\text{FS}_a}$ corresponds to the average over the respective Fermi surface. $|| \hat{M} ||^2 = \text{Tr}[\hat{M}\hat{M}^\dagger]$ corresponds to the Frobenius norm of the matrix $\hat{M}$. The superconducting fitness matrices are defined as
\begin{eqnarray}\label{DefFA}
\hat{F}_{A,C}(\bk,s) (i\hat{\sigma}_2) =\tilde{ H}_0(\bk,s) \tilde{\Delta}(\bk) \pm \tilde{\Delta}(\bk) \tilde{H}_0^*(-\bk,s),
\end{eqnarray} 
where $\tilde{H}_0(\bk,s) = [\hat{H}_0(\bk,s) - \tilde{h}_{00}(\bk,s)\hat{\sigma}_0\otimes \hat{\tau}_0]/|\tilde{h}(\bk,s)|$, with $|\tilde{h}(\bk,s)|^2 = \sum_{(a,b) \neq (0,0)} |\tilde{h}_{ab}(\bk,s)|^2$, is the normalized normal state Hamiltonian, and $\tilde{\Delta}(\bk) = \hat{\Delta}(\bk)/d_0$ is the normalized gap matrix.  The index $C$  ($A$) in $\hat{F}_{C(A)}(\bk,s) $ corresponds to the (anti-)commutator nature of this quantity. A finite $\hat{F}_{A}(\bk,s) $ corresponds to the weight of intra-band pairing, guaranteeing a robust weak coupling instability. Conversely, a finite $\hat{F}_{C}(\bk,s) $ corresponds to the weight of inter-band pairing, which is detrimental to the superconducting instability and reduces the critical temperature. The evaluation of $\hat{F}_A(\bk)$ for the local order parameters in $E_g$ within the two-orbital models is summarized in Table \ref{Tab:FAC2}. 

For small strain, we can determine the evolution of the superconducting fitness functions using the evolution of the hopping amplitudes and the total DOS  evaluated by DFT calculations (see SM \cite{SM} for quantitative estimates). Given the proximity of the two Fermi surfaces at the XZ and YZ planes, we assume that the superconducting fitness functions for each Fermi surface are going to be approximately the same. We define
\begin{eqnarray}\label{Eq:Fs}
\langle ||\hat{F}_A(\bk,s)||^2\rangle  & \approx &\langle ||\hat{F}_A(\bk,0)||^2 \rangle (1 + F_1 s + F_2 s^2),
\end{eqnarray}
for a representative Fermi vector, and the total density of states
\begin{eqnarray}\label{Eq:Ns}
N(0,s)& \approx &N(0,0) (1 + N_1 s + N_2 s^2).
\end{eqnarray}
The coefficients $F_{1,2}$ and $N_{1,2}$ are summarized in Table \ref{Tab:Rates} for compressive strain along different directions.

\begin{table}[t]
\begin{center}
\def\arraystretch{1.25}
\begin{tabular}{|c|c|c|c|c|c|}
\cline{2-6}
 \multicolumn{1}{c|}{} & $(3,0)_p$ & $(2,1)_p$ & $(2,2)_p$& $(2,3)_p$& $(1,0)_p$ \\
\hline
$[2,3]_p$  &0 & 0 & 0 & 1 & 0 \\ \hline
$[1,0]_p$  & 0 & 0 & 0 & 0 & 1 \\ \hline
\end{tabular}
\end{center}
\caption{ Superconducting fitness analysis for the effective two-orbital models. Each line corresponds to an order parameter with matrix structure $[a,b]_p$. The numerical entries correspond to $|| \hat F_{A}(\bk,s)||^2 = 4 \sum_{cd} (\text{table entry})\, |\tilde{h}_{cd}(\bk,s)|^2/|\tilde{h}(\bk,s)|^2$, for each term $(c,d)_p$ in the normal-state Hamiltonian. The corresponding table for $|| \hat F_{C}(\bk,s)||^2$ is obtained by exchanging $0 \leftrightarrow 1$.}
\label{Tab:FAC2}
\end{table}

\begin{center}
\begin{figure}[t]
\includegraphics[width=1.\linewidth, keepaspectratio]{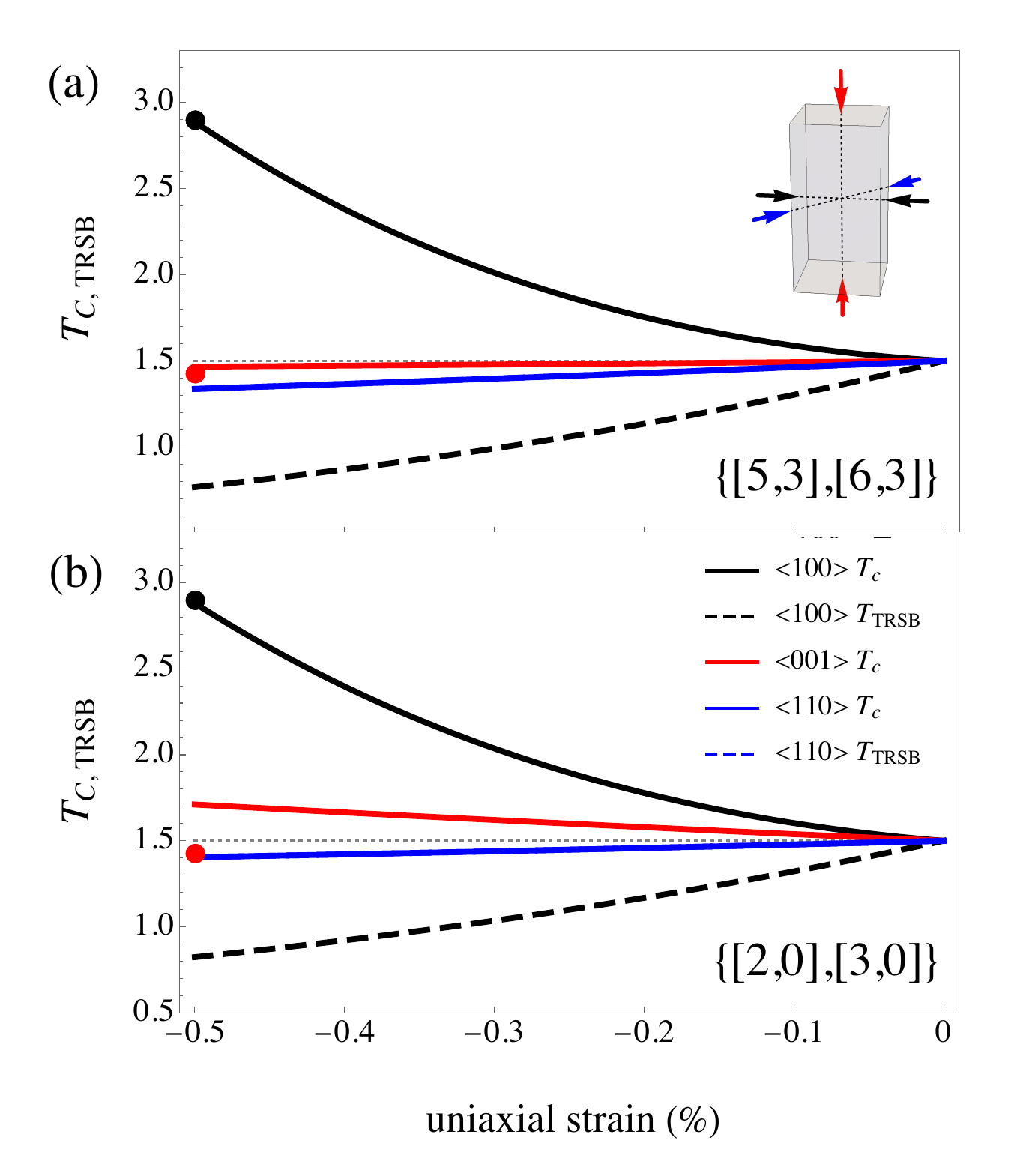}
\caption{Evolution of the critical temperatures under compressive uniaxial strain along the $\langle 100\rangle$, $\langle 001\rangle$ and $\langle 110\rangle$ directions. (a) Order parameter $\{[5,3],[6,3]\}$ with $g=5.8$. (b) Order parameter $\{[3,0],-[2,0]\}$ with $g=5$. The horizontal dashed line is a guide to the eye corresponding to the unstrained critical temperature. The black (red) circles at $s=-0.5\%$ correspond to the experimental value for strain along the $\langle 100\rangle$ ($\langle001\rangle$) direction. For both plots we set the representative $k_F = 2.67$ to be between the $\beta$ and $\gamma$ Fermi surfaces.}
\label{Fig:Tcs}
\end{figure}
\end{center}

In order to get a simpler form for the evolution of the critical temperature with strain, we use the relation $ ||\hat F_{A}(\bk)||^2+ ||\hat F_{C}(\bk)||^2 = 4$ and write
\begin{align}
\alpha(s) &\approx \alpha(0) (1+N_1 s + N_2 s^2)( 1+  F_1 s + F_2 s^2),\\ 
\delta(s)
&\approx \delta(0)(1+N_1 s + N_2 s^2)
\left[ 1- A ( F_1 s + F_2 s^2)\right],
\end{align}
where $A= \langle ||\hat{F}_A(0)||^2/q^2\rangle_{\text{FS}a}/ \langle ||\hat{F}_C(0)||^2/q^2\rangle_{\text{FS}a}$, with an implicit $\bk$ dependence. For the $[2,3]_{XZ/YZ}$ order parameters, we estimate $A \approx 10^{-5}$, and for the $[1,0]_{XZ/YZ}$ order parameters $A =0$ (see quantitative discussion in the SM \cite{SM}). The small value of $A$ allows us to neglect the dependence of $\delta(s)$ on strain through the $F_{1,2}$ coefficients. Furthermore, within the assumption that $1 \gg2|v| \delta(0)$, we take $\delta(s)\approx \delta(0)$, such that the dependence of the critical temperature on strain is carried only by the fitness function $\alpha(s)$. Within these considerations, the closed form equation for the evolution of $T_c$ can be cast as
\begin{eqnarray}
\frac{T_c (s)}{T_c(0)} &\approx& \exp{\left[g\left(1-\frac{\alpha(0)}{\alpha(s)} \right)\right]},
\end{eqnarray}
where $g=\left[1/(2|v|) + \delta(0)\right]/\alpha(0)$, the only free parameter in this analysis, is chosen such that our results are in good agreement with the experimentally observed value of  $T_c(s=-0.5\%) \approx 2.9$K for strain along the $\langle 100\rangle$  direction \cite{Steppke:2017}. Our results for strain along different directions are summarized in Fig. \ref{Fig:Tcs}. Note that for strain along the $\langle 100\rangle$ direction, the transition temperatures for $[2,3]_{XZ}$ and $[2,3]_{YZ}$ evolve differently, indicating the splitting of the two originally degenerate components $\{[5,3],[6,3]\}$ in $E_g$. We associate the higher temperature with $T_c$ and the lower temperature with $T_{\text{TRSB}}$. As expected, for strain along the $\langle 001\rangle$ direction $T_c$ and $T_{\text{TRSB}}$ are the same as there is no symmetry breaking. Finally, for strain along the $\langle 110 \rangle$ direction there is symmetry breaking, but its effect cannot be captured by our approach considering only the $XZ$ and $YZ$ planes since these planes remain equivalent (see further discussion in the SM \cite{SM}).

In order to qualitatively understand the evolution of the critical temperatures as a function of strain along different directions, we summarize here some of the properties of the parameters in the normal state Hamiltonian along the $XZ$ plane. The dominant term within $(3,0)_{XZ}$ is associated with the imbalance in the intra-orbital nearest-neighbour hopping amplitudes for the $d_{xz}$ and $d_{xy}$ orbitals. $(2,1)_{XZ}$ is associated with the correlation-enhanced effective atomic SOC, taken as $\eta \approx 0.2$ eV \cite{Tamai:2019}. $(2,3)_{XZ}$ is associated with momentum-dependent SOC in the $B_{2g}$ irrep. As an estimate, here we take its value to be equal to one tenth the hopping amplitude with the same inter-orbital structure \cite{Suh:2020}, $ t_{xy/xz}^{\text{SOC}}(\hat{d}_d) \approx  t_{xy/xz}(\hat{d}_d)/10 \approx 0.68$ meV. By symmetry, $(2,2)_{XZ}$ and $(1,0)_{XZ}$ are zero along the $XZ$ plane. We then focus on the evolution of the terms $(3,0)_{XZ}$ and $(2,3)_{XZ}$ with strain. The analysis for the $YZ$ plane leads to a similar conclusion. The evolution of the hopping amplitudes and DOS as a function of strain is determined by DFT calculations, performed without SOC. The correlation-enhanced effective atomic SOC is taken by its agreement with the experimental Fermi surfaces \cite{Tamai:2019} and was verified not to change significantly with strain (this result will be published elsewhere).

For compressive strain along the $\langle 100 \rangle$ direction, we see from Table \ref{Tab:Rates} that the dominant coefficients defining the evolution of the critical temperature with strain are associated with the evolution of the DOS ($|N_{1,2}|>|F_{1,2}|$). The effect of the fitness measure $|| \hat F_{A}(\bk,s)||^2$ is simply to slow down the evolution of the critical temperatures, as $F_{1,2}$ and $N_{1,2}$ have opposite sign. To note here is that at the linear level the evolution of the DOS, and therefore of the critical temperatures, is already rather asymmetric. Furthermore, for the $[5,3]$ component there is a very large quadratic contribution to the evolution of the DOS due to the proximity to a van Hove singularity, which reflects on the evolution of the critical temperature, as shown in Fig. \ref{Fig:Tcs} (a) (black curves). Given these coefficients obtained for low strain, we would expect an upturn of $T_{\text{TRSB}}$ for $s=-1.7\%$. While this is a relatively high strain value, the upturn can be accessed in experiments by deviations from linear behavior at low strains.

Compressive strain along the $\langle 001 \rangle$ direction does not reduce the symmetry group, and as a consequence there is no splitting of the superconducting transition temperatures. For this strain direction, we find $|F_1|>|N_1|$, what indicates that the behavior of the critical temperature as a function of strain is dominated by the evolution of the superconducting fitness measure $|| \hat F_{A}(\bk,s)||^2$. This is in contrast to the discussion above for strain along the $\langle 100 \rangle$, with the evolution of the critical temperature dominated by the DOS. In order to understand the evolution of the fitness measure under compressive strain, we can look more carefully at how each term in the normal state Hamiltonian evolves with strain. The superconducting fitness analysis in Table \ref{Tab:FAC2} tells us that, for the order parameter $[2,3]_{XZ}$, the normal state term $(3,0)_{XZ}$ contributes to a finite $\hat{F}_C(\bk)$ and $(2,3)_{XZ}$ contributes to a finite $\hat{F}_A(\bk)$. For compressive strain along the $\langle 001 \rangle$ direction, the magnitude of $(3,0)_{XZ} \propto t_{xz}(\hat{x})-t_{xy}(\hat{x})$ [here $t_{\gamma}(\hat{n})$ stands for the intra-orbital hopping amplitude for orbital $\gamma$ along the $\hat{n}$ direction] is enhanced, leading to an increase of $\hat{F}_C(\bk)$, and $(2,3)_{XZ}$ is reduced, leading to a reduction in $\hat{F}_A(\bk)$, such that the coefficient $F_1$ is negative. Interestingly, at the quadratic level the evolution of the critical temperature is dominated by the DOS, as $|N_2|>|F_2|$ and we predict an upturn of the critical temperature for strain values of approximately $- 2.2\%$. The results for the $\langle 001 \rangle$ direction are shown in Fig. \ref{Fig:Tcs}  (a) (red curve).

We also consider the behavior of the critical temperatures under compressive strain along the $\langle 110 \rangle$ direction. Strain along this direction reduces the point group symmetry, but as we are modelling only the $XZ$ and $YZ$ planes, we cannot capture the symmetry breaking and our results show the same strain evolution for both critical temperatures. In this case the behavior is dominated by the evolution of the DOS, this time in agreement with the evolution of the superconducting fitness function, as $F_{1,2}$ and $N_{1,2}$ have the same sign. We observe an overall reduction of the critical temperature for both components. These results are summarized in Fig. \ref{Fig:Tcs} (a) (blue curve). For this direction an upturn of the critical temperatures would also be expected for $s \approx - 2.9\%$. In a more complete treatment with a three dimensional Fermi surface, we expect to see a splitting of the critical temperatures away from the line we currently obtained. This suggests that if the splitting is smaller than the absolute reduction of the critical temperatures we have observed here, both $T_c$ and $T_{\text{TRSB}}$ could be suppressed as a function of compressive strain, as preliminary results indicate \cite{Henning:2021}.

Concerning the other two component order parameter in $E_g$, $\{[3,0],-[2,0]\}$, the strain dependence of $\hat{F}_A(\bk)$ is defined by the evolution of $\tilde{h}_{10}(\bk)$, but, by symmetry, this term is zero along the $XZ$ and $YZ$ planes. As consequence, within our approach, the evolution of the critical temperature is controlled only by the evolution of the DOS through the $N_{1,2}$ coefficients in Table \ref{Tab:Rates}. These results are summarized in Fig. \ref{Fig:Tcs} (b). Note that even though the behavior for strain along the $\langle100\rangle$ direction is similar to the $\{[5,3],[6,3]\}$ scenario (black curves), now compressive strain along the $\langle 001\rangle$ direction leads to an enhancement  of $T_c$ (red curve), while strain along the $\langle 110 \rangle$ reduces the critical temperature (blue curve). 

Here we would like to highlight that the reduction of the critical temperature for strain along the $\langle 001 \rangle$ direction seems to be a unique feature for the order parameter in $E_g$ dominated by $\{[5,3],[6,3]\}$, associated with the recently proposed OAST superconducting state. From a naive weak-coupling scenario, we would expect an increase in $T_c$ based on the enhanced DOS under strain, as recently discussed in Ref. \cite{Jerzembeck:2021}.

In conclusion, this work proposes a  microscopic perspective for the understanding of the effects of strain in superconductors with multiple degrees of freedom based on the superconducting fitness analysis. This construction is particularly useful in order to go beyond the predictions of effective Ginzburg-Landau theories. Assuming an OAST superconducting state with $E_g$ symmetry, we are able to account for the evolution of the critical temperatures under strain along different directions with a single free parameter $g$, associated with the strength of the interactions in this specific pairing channel. In particular, we find that compressive strain along the $\langle 100 \rangle$ direction can lead to an asymmetric splitting of $T_c$ and $T_{\text{TRSB}}$, and that compressive strain along the $\langle 001 \rangle$ direction reduces $T_c$, even though the DOS at the Fermi surface is enhanced. We also find that up to splitting, both $T_c$ and $T_{\text{TRSB}}$ are reduced under compressive strain along the $\langle 110 \rangle$ direction. These results indicate that the phenomenology of the OAST order parameter satisfies important constraints imposed by recent experiments. We believe that this kind of analysis can be extremely useful for  understanding the behavior of other complex superconductors under strain and to ultimately determine the symmetry of the superconducting order parameter in these materials.

\begin{widetext}

\begin{table}[ht]
\begin{center}
  \renewcommand\arraystretch{1.25}
    \begin{tabular}{| c | c | c | c | c | c |}
      \hline
Direction & SC Component & $N_1 (\%s)^{-1}$ & $N_2 (\%s)^{-2}$ & $ F_1 (\%s)^{-1}$ & $ F_2 (\%s)^{-2}$  \\ \hline \hline
      \multirow{2}{*}{$\langle 100 \rangle$} & $[5,3]$  &  $- 0.0928$ & $+0.4143$ & $+0.0333$ & $-0.0125$  \\ \cline{2-6}
       &$[6,3]$ & $+0.2536$ & $+0.0797$ & $-0.0091$ & $-0.0042$  \\ \hline
$\langle 001 \rangle$ & $\{[5,3],[6,3]\}$  & $-0.0503$ & $+0.0072$ & $+0.0589$ & $-0.0024$  \\ \hline
$\langle 110 \rangle$ & $\{[5,3],[6,3]\}$   & $+0.0296$ & $+ 0.0074$ & $+0.0129$ & $-0.0004$ \\\hline
    \end{tabular}
\end{center}
\caption{Coefficients for the evaluation of the evolution of the critical temperature as a function of strain for the  $\{[5,3],[6,3]\}$ order parameter  inferred from the analysis of the two-orbital models along the $XZ$ and $YZ$ planes. $N_1$ and $N_2$: Coefficients determining the evolution of the density of states at the Fermi energy. $F_1$ and $F_2$: Coefficients determining the evolution of $||\hat{F}_A(\bk,s)||^2$ (as in Eq. \ref{Eq:Fs}), obtained by selecting a representative $k_F = 2.67$ to be between the $\beta$ and $\gamma$ Fermi surfaces. Note that compressive strain correspond to negative values.}
\label{Tab:Rates}
\end{table}

\end{widetext}

\clearpage
\begin{acknowledgments}
The authors would like to thank S. Ghosh, V. Grinenko, C. Hicks, F. Jerzembeck, and H.-H. Klauss for interesting discussions. The Flatiron Institute is a division of the Simons Foundation. C.T. acknowledges support by the DFG through the Collaborative Research Center SFB 1143, project A04, the Research Training Group GRK 1621, and the Cluster of Excellence on Complexity and Topology in Quantum Matter ct.qmat (EXC 2147). A.R. acknowledges the financial support of the Swiss National Science Foundation through an Ambizione Grant No. 186043.
\end{acknowledgments}


\bibliography{SROStrain.V12}{}

\begin{thebibliography}{51}%
\makeatletter
\providecommand \@ifxundefined [1]{%
 \@ifx{#1\undefined}
}%
\providecommand \@ifnum [1]{%
 \ifnum #1\expandafter \@firstoftwo
 \else \expandafter \@secondoftwo
 \fi
}%
\providecommand \@ifx [1]{%
 \ifx #1\expandafter \@firstoftwo
 \else \expandafter \@secondoftwo
 \fi
}%
\providecommand \natexlab [1]{#1}%
\providecommand \enquote  [1]{``#1''}%
\providecommand \bibnamefont  [1]{#1}%
\providecommand \bibfnamefont [1]{#1}%
\providecommand \citenamefont [1]{#1}%
\providecommand \href@noop [0]{\@secondoftwo}%
\providecommand \href [0]{\begingroup \@sanitize@url \@href}%
\providecommand \@href[1]{\@@startlink{#1}\@@href}%
\providecommand \@@href[1]{\endgroup#1\@@endlink}%
\providecommand \@sanitize@url [0]{\catcode `\\12\catcode `\$12\catcode
  `\&12\catcode `\#12\catcode `\^12\catcode `\_12\catcode `\%12\relax}%
\providecommand \@@startlink[1]{}%
\providecommand \@@endlink[0]{}%
\providecommand \url  [0]{\begingroup\@sanitize@url \@url }%
\providecommand \@url [1]{\endgroup\@href {#1}{\urlprefix }}%
\providecommand \urlprefix  [0]{URL }%
\providecommand \Eprint [0]{\href }%
\providecommand \doibase [0]{http://dx.doi.org/}%
\providecommand \selectlanguage [0]{\@gobble}%
\providecommand \bibinfo  [0]{\@secondoftwo}%
\providecommand \bibfield  [0]{\@secondoftwo}%
\providecommand \translation [1]{[#1]}%
\providecommand \BibitemOpen [0]{}%
\providecommand \bibitemStop [0]{}%
\providecommand \bibitemNoStop [0]{.\EOS\space}%
\providecommand \EOS [0]{\spacefactor3000\relax}%
\providecommand \BibitemShut  [1]{\csname bibitem#1\endcsname}%
\let\auto@bib@innerbib\@empty
\bibitem [{\citenamefont {Hicks}\ \emph
  {et~al.}(2014{\natexlab{a}})\citenamefont {Hicks}, \citenamefont {Barber},
  \citenamefont {Edkins}, \citenamefont {Brodsky},\ and\ \citenamefont
  {Mackenzie}}]{Hicks:2014}%
  \BibitemOpen
  \bibfield  {author} {\bibinfo {author} {\bibfnamefont {Clifford~W.}\
  \bibnamefont {Hicks}}, \bibinfo {author} {\bibfnamefont {Mark~E.}\
  \bibnamefont {Barber}}, \bibinfo {author} {\bibfnamefont {Stephen~D.}\
  \bibnamefont {Edkins}}, \bibinfo {author} {\bibfnamefont {Daniel~O.}\
  \bibnamefont {Brodsky}}, \ and\ \bibinfo {author} {\bibfnamefont {Andrew~P.}\
  \bibnamefont {Mackenzie}},\ }\bibfield  {title} {\enquote {\bibinfo {title}
  {Piezoelectric-based apparatus for strain tuning},}\ }\href {\doibase
  10.1063/1.4881611} {\bibfield  {journal} {\bibinfo  {journal} {Review of
  Scientific Instruments}\ }\textbf {\bibinfo {volume} {85}},\ \bibinfo {pages}
  {065003} (\bibinfo {year} {2014}{\natexlab{a}})}\BibitemShut {NoStop}%
\bibitem [{\citenamefont {Brodsky}\ \emph {et~al.}(2017)\citenamefont
  {Brodsky}, \citenamefont {Barber}, \citenamefont {Bruin}, \citenamefont
  {Borzi}, \citenamefont {Grigera}, \citenamefont {Perry}, \citenamefont
  {Mackenzie},\ and\ \citenamefont {Hicks}}]{Brodskye:2017}%
  \BibitemOpen
  \bibfield  {author} {\bibinfo {author} {\bibfnamefont {Daniel~O.}\
  \bibnamefont {Brodsky}}, \bibinfo {author} {\bibfnamefont {Mark~E.}\
  \bibnamefont {Barber}}, \bibinfo {author} {\bibfnamefont {Jan A.~N.}\
  \bibnamefont {Bruin}}, \bibinfo {author} {\bibfnamefont {Rodolfo~A.}\
  \bibnamefont {Borzi}}, \bibinfo {author} {\bibfnamefont {Santiago~A.}\
  \bibnamefont {Grigera}}, \bibinfo {author} {\bibfnamefont {Robin~S.}\
  \bibnamefont {Perry}}, \bibinfo {author} {\bibfnamefont {Andrew~P.}\
  \bibnamefont {Mackenzie}}, \ and\ \bibinfo {author} {\bibfnamefont
  {Clifford~W.}\ \bibnamefont {Hicks}},\ }\bibfield  {title} {\enquote
  {\bibinfo {title} {Strain and vector magnetic field tuning of the anomalous
  phase in {S}r$_3${R}u$_2${O}$_7$},}\ }\href {\doibase 10.1126/sciadv.1501804}
  {\bibfield  {journal} {\bibinfo  {journal} {Science Advances}\ }\textbf
  {\bibinfo {volume} {3}},\ \bibinfo {pages} {e1501804} (\bibinfo {year}
  {2017})}\BibitemShut {NoStop}%
\bibitem [{\citenamefont {Park}\ \emph {et~al.}(2018)\citenamefont {Park},
  \citenamefont {Sakai}, \citenamefont {Erten}, \citenamefont {Mackenzie},\
  and\ \citenamefont {Hicks}}]{Park2:2018}%
  \BibitemOpen
  \bibfield  {author} {\bibinfo {author} {\bibfnamefont {Joonbum}\ \bibnamefont
  {Park}}, \bibinfo {author} {\bibfnamefont {Hideaki}\ \bibnamefont {Sakai}},
  \bibinfo {author} {\bibfnamefont {Onur}\ \bibnamefont {Erten}}, \bibinfo
  {author} {\bibfnamefont {Andrew~P.}\ \bibnamefont {Mackenzie}}, \ and\
  \bibinfo {author} {\bibfnamefont {Clifford~W.}\ \bibnamefont {Hicks}},\
  }\bibfield  {title} {\enquote {\bibinfo {title} {Effect of applied
  orthorhombic lattice distortion on the antiferromagnetic phase of
  {C}e{A}u{S}b$_{2}$},}\ }\href {\doibase 10.1103/PhysRevB.97.024411}
  {\bibfield  {journal} {\bibinfo  {journal} {Phys. Rev. B}\ }\textbf {\bibinfo
  {volume} {97}},\ \bibinfo {pages} {024411} (\bibinfo {year}
  {2018})}\BibitemShut {NoStop}%
\bibitem [{\citenamefont {B\"ohmer}\ \emph {et~al.}(2017)\citenamefont
  {B\"ohmer}, \citenamefont {Sapkota}, \citenamefont {Kreyssig}, \citenamefont
  {Bud'ko}, \citenamefont {Drachuck}, \citenamefont {Saunders}, \citenamefont
  {Goldman},\ and\ \citenamefont {Canfield}}]{Bohmer:2017}%
  \BibitemOpen
  \bibfield  {author} {\bibinfo {author} {\bibfnamefont {A.~E.}\ \bibnamefont
  {B\"ohmer}}, \bibinfo {author} {\bibfnamefont {A.}~\bibnamefont {Sapkota}},
  \bibinfo {author} {\bibfnamefont {A.}~\bibnamefont {Kreyssig}}, \bibinfo
  {author} {\bibfnamefont {S.~L.}\ \bibnamefont {Bud'ko}}, \bibinfo {author}
  {\bibfnamefont {G.}~\bibnamefont {Drachuck}}, \bibinfo {author}
  {\bibfnamefont {S.~M.}\ \bibnamefont {Saunders}}, \bibinfo {author}
  {\bibfnamefont {A.~I.}\ \bibnamefont {Goldman}}, \ and\ \bibinfo {author}
  {\bibfnamefont {P.~C.}\ \bibnamefont {Canfield}},\ }\bibfield  {title}
  {\enquote {\bibinfo {title} {Effect of biaxial strain on the phase
  transitions of {C}a({F}e$_{1-x}${C}o$_x$){A}s$_2$},}\ }\href {\doibase
  10.1103/PhysRevLett.118.107002} {\bibfield  {journal} {\bibinfo  {journal}
  {Phys. Rev. Lett.}\ }\textbf {\bibinfo {volume} {118}},\ \bibinfo {pages}
  {107002} (\bibinfo {year} {2017})}\BibitemShut {NoStop}%
\bibitem [{\citenamefont {Kim}\ \emph {et~al.}(2018)\citenamefont {Kim},
  \citenamefont {Souliou}, \citenamefont {Barber}, \citenamefont {Lefran{\c
  c}ois}, \citenamefont {Minola}, \citenamefont {Tortora}, \citenamefont
  {Heid}, \citenamefont {Nandi}, \citenamefont {Borzi}, \citenamefont
  {Garbarino}, \citenamefont {Bosak}, \citenamefont {Porras}, \citenamefont
  {Loew}, \citenamefont {K{\"o}nig}, \citenamefont {Moll}, \citenamefont
  {Mackenzie}, \citenamefont {Keimer}, \citenamefont {Hicks},\ and\
  \citenamefont {Le~Tacon}}]{Kim:2018}%
  \BibitemOpen
  \bibfield  {author} {\bibinfo {author} {\bibfnamefont {H.-H.}\ \bibnamefont
  {Kim}}, \bibinfo {author} {\bibfnamefont {S.~M.}\ \bibnamefont {Souliou}},
  \bibinfo {author} {\bibfnamefont {M.~E.}\ \bibnamefont {Barber}}, \bibinfo
  {author} {\bibfnamefont {E.}~\bibnamefont {Lefran{\c c}ois}}, \bibinfo
  {author} {\bibfnamefont {M.}~\bibnamefont {Minola}}, \bibinfo {author}
  {\bibfnamefont {M.}~\bibnamefont {Tortora}}, \bibinfo {author} {\bibfnamefont
  {R.}~\bibnamefont {Heid}}, \bibinfo {author} {\bibfnamefont {N.}~\bibnamefont
  {Nandi}}, \bibinfo {author} {\bibfnamefont {R.~A.}\ \bibnamefont {Borzi}},
  \bibinfo {author} {\bibfnamefont {G.}~\bibnamefont {Garbarino}}, \bibinfo
  {author} {\bibfnamefont {A.}~\bibnamefont {Bosak}}, \bibinfo {author}
  {\bibfnamefont {J.}~\bibnamefont {Porras}}, \bibinfo {author} {\bibfnamefont
  {T.}~\bibnamefont {Loew}}, \bibinfo {author} {\bibfnamefont {M.}~\bibnamefont
  {K{\"o}nig}}, \bibinfo {author} {\bibfnamefont {P.~J.~W.}\ \bibnamefont
  {Moll}}, \bibinfo {author} {\bibfnamefont {A.~P.}\ \bibnamefont {Mackenzie}},
  \bibinfo {author} {\bibfnamefont {B.}~\bibnamefont {Keimer}}, \bibinfo
  {author} {\bibfnamefont {C.~W.}\ \bibnamefont {Hicks}}, \ and\ \bibinfo
  {author} {\bibfnamefont {M.}~\bibnamefont {Le~Tacon}},\ }\bibfield  {title}
  {\enquote {\bibinfo {title} {Uniaxial pressure control of competing orders in
  a high-temperature superconductor},}\ }\href {\doibase
  10.1126/science.aat4708} {\bibfield  {journal} {\bibinfo  {journal}
  {Science}\ }\textbf {\bibinfo {volume} {362}},\ \bibinfo {pages} {1040--1044}
  (\bibinfo {year} {2018})}\BibitemShut {NoStop}%
\bibitem [{\citenamefont {Sun}\ \emph {et~al.}(2019)\citenamefont {Sun},
  \citenamefont {Sokolov}, \citenamefont {Bartlett}, \citenamefont
  {Sannigrahi}, \citenamefont {Khim}, \citenamefont {Kushwaha}, \citenamefont
  {Khalyavin}, \citenamefont {Manuel}, \citenamefont {Gibbs}, \citenamefont
  {Takagi}, \citenamefont {Mackenzie},\ and\ \citenamefont {Hicks}}]{Sun:2019}%
  \BibitemOpen
  \bibfield  {author} {\bibinfo {author} {\bibfnamefont {Dan}\ \bibnamefont
  {Sun}}, \bibinfo {author} {\bibfnamefont {Dmitry~A.}\ \bibnamefont
  {Sokolov}}, \bibinfo {author} {\bibfnamefont {Jack~M.}\ \bibnamefont
  {Bartlett}}, \bibinfo {author} {\bibfnamefont {Jhuma}\ \bibnamefont
  {Sannigrahi}}, \bibinfo {author} {\bibfnamefont {Seunghyun}\ \bibnamefont
  {Khim}}, \bibinfo {author} {\bibfnamefont {Pallavi}\ \bibnamefont
  {Kushwaha}}, \bibinfo {author} {\bibfnamefont {Dmitry~D.}\ \bibnamefont
  {Khalyavin}}, \bibinfo {author} {\bibfnamefont {Pascal}\ \bibnamefont
  {Manuel}}, \bibinfo {author} {\bibfnamefont {Alexandra~S.}\ \bibnamefont
  {Gibbs}}, \bibinfo {author} {\bibfnamefont {Hidenori}\ \bibnamefont
  {Takagi}}, \bibinfo {author} {\bibfnamefont {Andrew~P.}\ \bibnamefont
  {Mackenzie}}, \ and\ \bibinfo {author} {\bibfnamefont {Clifford~W.}\
  \bibnamefont {Hicks}},\ }\bibfield  {title} {\enquote {\bibinfo {title}
  {Magnetic frustration and spontaneous rotational symmetry breaking in
  {P}d{C}r{O}$_{2}$},}\ }\href {\doibase 10.1103/PhysRevB.100.094414}
  {\bibfield  {journal} {\bibinfo  {journal} {Phys. Rev. B}\ }\textbf {\bibinfo
  {volume} {100}},\ \bibinfo {pages} {094414} (\bibinfo {year}
  {2019})}\BibitemShut {NoStop}%
\bibitem [{\citenamefont {Pfau}\ \emph {et~al.}(2019)\citenamefont {Pfau},
  \citenamefont {Rotundu}, \citenamefont {Palmstrom}, \citenamefont {Chen},
  \citenamefont {Hashimoto}, \citenamefont {Lu}, \citenamefont {Kemper},
  \citenamefont {Fisher},\ and\ \citenamefont {Shen}}]{Pfau:2019}%
  \BibitemOpen
  \bibfield  {author} {\bibinfo {author} {\bibfnamefont {H.}~\bibnamefont
  {Pfau}}, \bibinfo {author} {\bibfnamefont {C.~R.}\ \bibnamefont {Rotundu}},
  \bibinfo {author} {\bibfnamefont {J.~C.}\ \bibnamefont {Palmstrom}}, \bibinfo
  {author} {\bibfnamefont {S.~D.}\ \bibnamefont {Chen}}, \bibinfo {author}
  {\bibfnamefont {M.}~\bibnamefont {Hashimoto}}, \bibinfo {author}
  {\bibfnamefont {D.}~\bibnamefont {Lu}}, \bibinfo {author} {\bibfnamefont
  {A.~F.}\ \bibnamefont {Kemper}}, \bibinfo {author} {\bibfnamefont {I.~R.}\
  \bibnamefont {Fisher}}, \ and\ \bibinfo {author} {\bibfnamefont {Z.-X.}\
  \bibnamefont {Shen}},\ }\bibfield  {title} {\enquote {\bibinfo {title}
  {Detailed band structure of twinned and detwinned {B}a{F}e$_{2}${A}s$_{2}$
  studied with angle-resolved photoemission spectroscopy},}\ }\href {\doibase
  10.1103/PhysRevB.99.035118} {\bibfield  {journal} {\bibinfo  {journal} {Phys.
  Rev. B}\ }\textbf {\bibinfo {volume} {99}},\ \bibinfo {pages} {035118}
  (\bibinfo {year} {2019})}\BibitemShut {NoStop}%
\bibitem [{\citenamefont {Hsu}\ \emph {et~al.}(2016)\citenamefont {Hsu},
  \citenamefont {Cho}, \citenamefont {Rebola}, \citenamefont {Burganov},
  \citenamefont {Adamo}, \citenamefont {Shen}, \citenamefont {Schlom},
  \citenamefont {Fennie},\ and\ \citenamefont {Kim}}]{Hsu:2016}%
  \BibitemOpen
  \bibfield  {author} {\bibinfo {author} {\bibfnamefont {Yi-Ting}\ \bibnamefont
  {Hsu}}, \bibinfo {author} {\bibfnamefont {Weejee}\ \bibnamefont {Cho}},
  \bibinfo {author} {\bibfnamefont {Alejandro~Federico}\ \bibnamefont
  {Rebola}}, \bibinfo {author} {\bibfnamefont {Bulat}\ \bibnamefont
  {Burganov}}, \bibinfo {author} {\bibfnamefont {Carolina}\ \bibnamefont
  {Adamo}}, \bibinfo {author} {\bibfnamefont {Kyle~M.}\ \bibnamefont {Shen}},
  \bibinfo {author} {\bibfnamefont {Darrell~G.}\ \bibnamefont {Schlom}},
  \bibinfo {author} {\bibfnamefont {Craig~J.}\ \bibnamefont {Fennie}}, \ and\
  \bibinfo {author} {\bibfnamefont {Eun-Ah}\ \bibnamefont {Kim}},\ }\bibfield
  {title} {\enquote {\bibinfo {title} {Manipulating superconductivity in
  ruthenates through fermi surface engineering},}\ }\href {\doibase
  10.1103/PhysRevB.94.045118} {\bibfield  {journal} {\bibinfo  {journal} {Phys.
  Rev. B}\ }\textbf {\bibinfo {volume} {94}},\ \bibinfo {pages} {045118}
  (\bibinfo {year} {2016})}\BibitemShut {NoStop}%
\bibitem [{\citenamefont {Mutch}\ \emph {et~al.}(2019)\citenamefont {Mutch},
  \citenamefont {Chen}, \citenamefont {Went}, \citenamefont {Qian},
  \citenamefont {Wilson}, \citenamefont {Andreev}, \citenamefont {Chen},\ and\
  \citenamefont {Chu}}]{Mutch:2019}%
  \BibitemOpen
  \bibfield  {author} {\bibinfo {author} {\bibfnamefont {Joshua}\ \bibnamefont
  {Mutch}}, \bibinfo {author} {\bibfnamefont {Wei-Chih}\ \bibnamefont {Chen}},
  \bibinfo {author} {\bibfnamefont {Preston}\ \bibnamefont {Went}}, \bibinfo
  {author} {\bibfnamefont {Tiema}\ \bibnamefont {Qian}}, \bibinfo {author}
  {\bibfnamefont {Ilham~Zaky}\ \bibnamefont {Wilson}}, \bibinfo {author}
  {\bibfnamefont {Anton}\ \bibnamefont {Andreev}}, \bibinfo {author}
  {\bibfnamefont {Cheng-Chien}\ \bibnamefont {Chen}}, \ and\ \bibinfo {author}
  {\bibfnamefont {Jiun-Haw}\ \bibnamefont {Chu}},\ }\bibfield  {title}
  {\enquote {\bibinfo {title} {Evidence for a strain-tuned topological phase
  transition in {Z}r{T}e$_5$},}\ }\href {\doibase 10.1126/sciadv.aav9771}
  {\bibfield  {journal} {\bibinfo  {journal} {Science Advances}\ }\textbf
  {\bibinfo {volume} {5}},\ \bibinfo {pages} {eaav9771} (\bibinfo {year}
  {2019})}\BibitemShut {NoStop}%
\bibitem [{\citenamefont {Mackenzie}\ \emph {et~al.}(2017)\citenamefont
  {Mackenzie}, \citenamefont {Scaffidi}, \citenamefont {Hicks},\ and\
  \citenamefont {Maeno}}]{Mackenzie:2017}%
  \BibitemOpen
  \bibfield  {author} {\bibinfo {author} {\bibfnamefont {Andrew~P.}\
  \bibnamefont {Mackenzie}}, \bibinfo {author} {\bibfnamefont {Thomas}\
  \bibnamefont {Scaffidi}}, \bibinfo {author} {\bibfnamefont {Clifford~W.}\
  \bibnamefont {Hicks}}, \ and\ \bibinfo {author} {\bibfnamefont {Yoshiteru}\
  \bibnamefont {Maeno}},\ }\bibfield  {title} {\enquote {\bibinfo {title} {Even
  odder after twenty-three years: the superconducting order parameter puzzle of
  {S}r$_2${R}u{O}$_4$},}\ }\href {\doibase 10.1038/s41535-017-0045-4}
  {\bibfield  {journal} {\bibinfo  {journal} {npj Quantum Materials}\ }\textbf
  {\bibinfo {volume} {2}},\ \bibinfo {pages} {40} (\bibinfo {year}
  {2017})}\BibitemShut {NoStop}%
\bibitem [{\citenamefont {Watson}\ \emph {et~al.}(2018)\citenamefont {Watson},
  \citenamefont {Gibbs}, \citenamefont {Mackenzie}, \citenamefont {Hicks},\
  and\ \citenamefont {Moler}}]{Watson:2018}%
  \BibitemOpen
  \bibfield  {author} {\bibinfo {author} {\bibfnamefont {Christopher~A.}\
  \bibnamefont {Watson}}, \bibinfo {author} {\bibfnamefont {Alexandra~S.}\
  \bibnamefont {Gibbs}}, \bibinfo {author} {\bibfnamefont {Andrew~P.}\
  \bibnamefont {Mackenzie}}, \bibinfo {author} {\bibfnamefont {Clifford~W.}\
  \bibnamefont {Hicks}}, \ and\ \bibinfo {author} {\bibfnamefont {Kathryn~A.}\
  \bibnamefont {Moler}},\ }\bibfield  {title} {\enquote {\bibinfo {title}
  {Micron-scale measurements of low anisotropic strain response of local
  ${T}_{c}$ in {S}r$_2${R}u{O}$_4$},}\ }\href {\doibase
  10.1103/PhysRevB.98.094521} {\bibfield  {journal} {\bibinfo  {journal} {Phys.
  Rev. B}\ }\textbf {\bibinfo {volume} {98}},\ \bibinfo {pages} {094521}
  (\bibinfo {year} {2018})}\BibitemShut {NoStop}%
\bibitem [{\citenamefont {Steppke}\ \emph {et~al.}(2017)\citenamefont
  {Steppke}, \citenamefont {Zhao}, \citenamefont {Barber}, \citenamefont
  {Scaffidi}, \citenamefont {Jerzembeck}, \citenamefont {Rosner}, \citenamefont
  {Gibbs}, \citenamefont {Maeno}, \citenamefont {Simon}, \citenamefont
  {Mackenzie},\ and\ \citenamefont {Hicks}}]{Steppke:2017}%
  \BibitemOpen
  \bibfield  {author} {\bibinfo {author} {\bibfnamefont {Alexander}\
  \bibnamefont {Steppke}}, \bibinfo {author} {\bibfnamefont {Lishan}\
  \bibnamefont {Zhao}}, \bibinfo {author} {\bibfnamefont {Mark~E.}\
  \bibnamefont {Barber}}, \bibinfo {author} {\bibfnamefont {Thomas}\
  \bibnamefont {Scaffidi}}, \bibinfo {author} {\bibfnamefont {Fabian}\
  \bibnamefont {Jerzembeck}}, \bibinfo {author} {\bibfnamefont {Helge}\
  \bibnamefont {Rosner}}, \bibinfo {author} {\bibfnamefont {Alexandra~S.}\
  \bibnamefont {Gibbs}}, \bibinfo {author} {\bibfnamefont {Yoshiteru}\
  \bibnamefont {Maeno}}, \bibinfo {author} {\bibfnamefont {Steven~H.}\
  \bibnamefont {Simon}}, \bibinfo {author} {\bibfnamefont {Andrew~P.}\
  \bibnamefont {Mackenzie}}, \ and\ \bibinfo {author} {\bibfnamefont
  {Clifford~W.}\ \bibnamefont {Hicks}},\ }\bibfield  {title} {\enquote
  {\bibinfo {title} {Strong peak in tc of {S}r$_2${R}u{O}$_4$ under uniaxial
  pressure},}\ }\href {\doibase 10.1126/science.aaf9398} {\bibfield  {journal}
  {\bibinfo  {journal} {Science}\ }\textbf {\bibinfo {volume} {355}},\ \bibinfo
  {pages} {eaaf9398} (\bibinfo {year} {2017})}\BibitemShut {NoStop}%
\bibitem [{\citenamefont {Barber}\ \emph {et~al.}(2018)\citenamefont {Barber},
  \citenamefont {Gibbs}, \citenamefont {Maeno}, \citenamefont {Mackenzie},\
  and\ \citenamefont {Hicks}}]{Barber:2018}%
  \BibitemOpen
  \bibfield  {author} {\bibinfo {author} {\bibfnamefont {M.~E.}\ \bibnamefont
  {Barber}}, \bibinfo {author} {\bibfnamefont {A.~S.}\ \bibnamefont {Gibbs}},
  \bibinfo {author} {\bibfnamefont {Y.}~\bibnamefont {Maeno}}, \bibinfo
  {author} {\bibfnamefont {A.~P.}\ \bibnamefont {Mackenzie}}, \ and\ \bibinfo
  {author} {\bibfnamefont {C.~W.}\ \bibnamefont {Hicks}},\ }\bibfield  {title}
  {\enquote {\bibinfo {title} {Resistivity in the vicinity of a van hove
  singularity: {S}r$_2${R}u{O}$_4$ under uniaxial pressure},}\ }\href {\doibase
  10.1103/PhysRevLett.120.076602} {\bibfield  {journal} {\bibinfo  {journal}
  {Phys. Rev. Lett.}\ }\textbf {\bibinfo {volume} {120}},\ \bibinfo {pages}
  {076602} (\bibinfo {year} {2018})}\BibitemShut {NoStop}%
\bibitem [{\citenamefont {Sunko}\ \emph {et~al.}(2019)\citenamefont {Sunko},
  \citenamefont {Morales}, \citenamefont {Marković}, \citenamefont {Barber},
  \citenamefont {Milosavljević}, \citenamefont {Mazzola}, \citenamefont
  {Sokolov}, \citenamefont {Kikugawa}, \citenamefont {Cacho}, \citenamefont
  {Dudin}, \citenamefont {Rosner}, \citenamefont {Hicks}, \citenamefont
  {King},\ and\ \citenamefont {Mackenzie}}]{Sunko:2019}%
  \BibitemOpen
  \bibfield  {author} {\bibinfo {author} {\bibfnamefont {Veronika}\
  \bibnamefont {Sunko}}, \bibinfo {author} {\bibfnamefont {Edgar~Abarca}\
  \bibnamefont {Morales}}, \bibinfo {author} {\bibfnamefont {Igor}\
  \bibnamefont {Marković}}, \bibinfo {author} {\bibfnamefont {Mark~E.}\
  \bibnamefont {Barber}}, \bibinfo {author} {\bibfnamefont {Dijana}\
  \bibnamefont {Milosavljević}}, \bibinfo {author} {\bibfnamefont {Federico}\
  \bibnamefont {Mazzola}}, \bibinfo {author} {\bibfnamefont {Dmitry~A.}\
  \bibnamefont {Sokolov}}, \bibinfo {author} {\bibfnamefont {Naoki}\
  \bibnamefont {Kikugawa}}, \bibinfo {author} {\bibfnamefont {Cephise}\
  \bibnamefont {Cacho}}, \bibinfo {author} {\bibfnamefont {Pavel}\ \bibnamefont
  {Dudin}}, \bibinfo {author} {\bibfnamefont {Helge}\ \bibnamefont {Rosner}},
  \bibinfo {author} {\bibfnamefont {Clifford~W.}\ \bibnamefont {Hicks}},
  \bibinfo {author} {\bibfnamefont {Philip D.~C.}\ \bibnamefont {King}}, \ and\
  \bibinfo {author} {\bibfnamefont {Andrew~P.}\ \bibnamefont {Mackenzie}},\
  }\bibfield  {title} {\enquote {\bibinfo {title} {Direct observation of a
  uniaxial stress-driven lifshitz transition in {S}r$_2${R}u{O}$_4$},}\ }\href
  {https://www.nature.com/articles/s41535-019-0185-9#citeas} {\bibfield
  {journal} {\bibinfo  {journal} {npj Quantum Materials}\ }\textbf {\bibinfo
  {volume} {4}},\ \bibinfo {pages} {46} (\bibinfo {year} {2019})}\BibitemShut
  {NoStop}%
\bibitem [{\citenamefont {{Pustogow}}\ \emph {et~al.}(2019)\citenamefont
  {{Pustogow}}, \citenamefont {{Luo}}, \citenamefont {{Chronister}},
  \citenamefont {{Su}}, \citenamefont {{Sokolov}}, \citenamefont
  {{Jerzembeck}}, \citenamefont {{Mackenzie}}, \citenamefont {{Hicks}},
  \citenamefont {{Kikugawa}}, \citenamefont {{Raghu}}, \citenamefont
  {{Bauer}},\ and\ \citenamefont {{Brown}}}]{Pustogow:2019}%
  \BibitemOpen
  \bibfield  {author} {\bibinfo {author} {\bibfnamefont {A.}~\bibnamefont
  {{Pustogow}}}, \bibinfo {author} {\bibfnamefont {Yongkang}\ \bibnamefont
  {{Luo}}}, \bibinfo {author} {\bibfnamefont {A.}~\bibnamefont {{Chronister}}},
  \bibinfo {author} {\bibfnamefont {Y.~S.}\ \bibnamefont {{Su}}}, \bibinfo
  {author} {\bibfnamefont {D.~A.}\ \bibnamefont {{Sokolov}}}, \bibinfo {author}
  {\bibfnamefont {F.}~\bibnamefont {{Jerzembeck}}}, \bibinfo {author}
  {\bibfnamefont {A.~P.}\ \bibnamefont {{Mackenzie}}}, \bibinfo {author}
  {\bibfnamefont {C.~W.}\ \bibnamefont {{Hicks}}}, \bibinfo {author}
  {\bibfnamefont {N.}~\bibnamefont {{Kikugawa}}}, \bibinfo {author}
  {\bibfnamefont {S.}~\bibnamefont {{Raghu}}}, \bibinfo {author} {\bibfnamefont
  {E.~D.}\ \bibnamefont {{Bauer}}}, \ and\ \bibinfo {author} {\bibfnamefont
  {S.~E.}\ \bibnamefont {{Brown}}},\ }\bibfield  {title} {\enquote {\bibinfo
  {title} {{Constraints on the superconducting order parameter in
  Sr$_{2}$RuO$_{4}$ from oxygen-17 nuclear magnetic resonance}},}\ }\href
  {\doibase 10.1038/s41586-019-1596-2} {\bibfield  {journal} {\bibinfo
  {journal} {\nat}\ }\textbf {\bibinfo {volume} {574}},\ \bibinfo {pages}
  {72--75} (\bibinfo {year} {2019})}\BibitemShut {NoStop}%
\bibitem [{\citenamefont {Grinenko}\ \emph {et~al.}(2021)\citenamefont
  {Grinenko}, \citenamefont {Ghosh}, \citenamefont {Sarkar}, \citenamefont
  {Orain}, \citenamefont {Nikitin}, \citenamefont {Elender}, \citenamefont
  {Das}, \citenamefont {Guguchia}, \citenamefont {Br{\"{u}}ckner},
  \citenamefont {Barber}, \citenamefont {Park}, \citenamefont {Kikugawa},
  \citenamefont {Sokolov}, \citenamefont {Bobowski}, \citenamefont {Miyoshi},
  \citenamefont {Maeno}, \citenamefont {Mackenzie}, \citenamefont {Luetkens},
  \citenamefont {Hicks},\ and\ \citenamefont {Klauss}}]{Grinenko:2021}%
  \BibitemOpen
  \bibfield  {author} {\bibinfo {author} {\bibfnamefont {Vadim}\ \bibnamefont
  {Grinenko}}, \bibinfo {author} {\bibfnamefont {Shreenanda}\ \bibnamefont
  {Ghosh}}, \bibinfo {author} {\bibfnamefont {Rajib}\ \bibnamefont {Sarkar}},
  \bibinfo {author} {\bibfnamefont {Jean-Christophe}\ \bibnamefont {Orain}},
  \bibinfo {author} {\bibfnamefont {Artem}\ \bibnamefont {Nikitin}}, \bibinfo
  {author} {\bibfnamefont {Matthias}\ \bibnamefont {Elender}}, \bibinfo
  {author} {\bibfnamefont {Debarchan}\ \bibnamefont {Das}}, \bibinfo {author}
  {\bibfnamefont {Zurab}\ \bibnamefont {Guguchia}}, \bibinfo {author}
  {\bibfnamefont {Felix}\ \bibnamefont {Br{\"{u}}ckner}}, \bibinfo {author}
  {\bibfnamefont {Mark~E}\ \bibnamefont {Barber}}, \bibinfo {author}
  {\bibfnamefont {Joonbum}\ \bibnamefont {Park}}, \bibinfo {author}
  {\bibfnamefont {Naoki}\ \bibnamefont {Kikugawa}}, \bibinfo {author}
  {\bibfnamefont {Dmitry~A}\ \bibnamefont {Sokolov}}, \bibinfo {author}
  {\bibfnamefont {Jake~S}\ \bibnamefont {Bobowski}}, \bibinfo {author}
  {\bibfnamefont {Takuto}\ \bibnamefont {Miyoshi}}, \bibinfo {author}
  {\bibfnamefont {Yoshiteru}\ \bibnamefont {Maeno}}, \bibinfo {author}
  {\bibfnamefont {Andrew~P}\ \bibnamefont {Mackenzie}}, \bibinfo {author}
  {\bibfnamefont {Hubertus}\ \bibnamefont {Luetkens}}, \bibinfo {author}
  {\bibfnamefont {Clifford~W}\ \bibnamefont {Hicks}}, \ and\ \bibinfo {author}
  {\bibfnamefont {Hans-Henning}\ \bibnamefont {Klauss}},\ }\bibfield  {title}
  {\enquote {\bibinfo {title} {{Split superconducting and time-reversal
  symmetry-breaking transitions in {S}r$_2${R}u{O}$_4$ under stress}},}\ }\href
  {\doibase 10.1038/s41567-021-01182-7} {\bibfield  {journal} {\bibinfo
  {journal} {Nature Physics}\ }\textbf {\bibinfo {volume} {17}},\ \bibinfo
  {pages} {748--754} (\bibinfo {year} {2021})}\BibitemShut {NoStop}%
\bibitem [{\citenamefont {Rice}\ and\ \citenamefont
  {Sigrist}(1995)}]{Rice:1995}%
  \BibitemOpen
  \bibfield  {author} {\bibinfo {author} {\bibfnamefont {T~M}\ \bibnamefont
  {Rice}}\ and\ \bibinfo {author} {\bibfnamefont {M}~\bibnamefont {Sigrist}},\
  }\bibfield  {title} {\enquote {\bibinfo {title} {{S}r$_2${R}u{O}$_4$: an
  electronic analogue of $^3${H}e?}}\ }\href {\doibase
  10.1088/0953-8984/7/47/002} {\bibfield  {journal} {\bibinfo  {journal}
  {Journal of Physics: Condensed Matter}\ }\textbf {\bibinfo {volume} {7}},\
  \bibinfo {pages} {L643--L648} (\bibinfo {year} {1995})}\BibitemShut {NoStop}%
\bibitem [{\citenamefont {Kallin}\ and\ \citenamefont
  {Berlinsky}(2009)}]{Kallin:2009}%
  \BibitemOpen
  \bibfield  {author} {\bibinfo {author} {\bibfnamefont {C}~\bibnamefont
  {Kallin}}\ and\ \bibinfo {author} {\bibfnamefont {A~J}\ \bibnamefont
  {Berlinsky}},\ }\bibfield  {title} {\enquote {\bibinfo {title} {Is
  {S}r$_2${R}u{O}$_4$ a chiral p-wave superconductor?}}\ }\href {\doibase
  10.1088/0953-8984/21/16/164210} {\bibfield  {journal} {\bibinfo  {journal}
  {Journal of Physics: Condensed Matter}\ }\textbf {\bibinfo {volume} {21}},\
  \bibinfo {pages} {164210} (\bibinfo {year} {2009})}\BibitemShut {NoStop}%
\bibitem [{\citenamefont {Kallin}(2012)}]{Kallin:2012}%
  \BibitemOpen
  \bibfield  {author} {\bibinfo {author} {\bibfnamefont {Catherine}\
  \bibnamefont {Kallin}},\ }\bibfield  {title} {\enquote {\bibinfo {title}
  {Chiral p-wave order in {S}r$_2${R}u{O}$_4$},}\ }\href {\doibase
  10.1088/0034-4885/75/4/042501} {\bibfield  {journal} {\bibinfo  {journal}
  {Reports on Progress in Physics}\ }\textbf {\bibinfo {volume} {75}},\
  \bibinfo {pages} {042501} (\bibinfo {year} {2012})}\BibitemShut {NoStop}%
\bibitem [{\citenamefont {{Kallin}}\ and\ \citenamefont
  {{Berlinsky}}(2016)}]{Kallin:2016}%
  \BibitemOpen
  \bibfield  {author} {\bibinfo {author} {\bibfnamefont {Catherine}\
  \bibnamefont {{Kallin}}}\ and\ \bibinfo {author} {\bibfnamefont {John}\
  \bibnamefont {{Berlinsky}}},\ }\bibfield  {title} {\enquote {\bibinfo {title}
  {{Chiral superconductors}},}\ }\href {\doibase 10.1088/0034-4885/79/5/054502}
  {\bibfield  {journal} {\bibinfo  {journal} {Reports on Progress in Physics}\
  }\textbf {\bibinfo {volume} {79}},\ \bibinfo {eid} {054502} (\bibinfo {year}
  {2016})},\ \Eprint {http://arxiv.org/abs/1512.01151} {arXiv:1512.01151
  [cond-mat.supr-con]} \BibitemShut {NoStop}%
\bibitem [{\citenamefont {Xia}\ \emph {et~al.}(2006)\citenamefont {Xia},
  \citenamefont {Maeno}, \citenamefont {Beyersdorf}, \citenamefont {Fejer},\
  and\ \citenamefont {Kapitulnik}}]{Xia:2006}%
  \BibitemOpen
  \bibfield  {author} {\bibinfo {author} {\bibfnamefont {Jing}\ \bibnamefont
  {Xia}}, \bibinfo {author} {\bibfnamefont {Yoshiteru}\ \bibnamefont {Maeno}},
  \bibinfo {author} {\bibfnamefont {Peter~T.}\ \bibnamefont {Beyersdorf}},
  \bibinfo {author} {\bibfnamefont {M.~M.}\ \bibnamefont {Fejer}}, \ and\
  \bibinfo {author} {\bibfnamefont {Aharon}\ \bibnamefont {Kapitulnik}},\
  }\bibfield  {title} {\enquote {\bibinfo {title} {High resolution polar kerr
  effect measurements of {S}r$_2${R}u{O}$_4$: Evidence for broken time-reversal
  symmetry in the superconducting state},}\ }\href {\doibase
  10.1103/PhysRevLett.97.167002} {\bibfield  {journal} {\bibinfo  {journal}
  {Phys. Rev. Lett.}\ }\textbf {\bibinfo {volume} {97}},\ \bibinfo {pages}
  {167002} (\bibinfo {year} {2006})}\BibitemShut {NoStop}%
\bibitem [{\citenamefont {Lupien}\ \emph {et~al.}(2001)\citenamefont {Lupien},
  \citenamefont {MacFarlane}, \citenamefont {Proust}, \citenamefont
  {Taillefer}, \citenamefont {Mao},\ and\ \citenamefont {Maeno}}]{Lupien:2001}%
  \BibitemOpen
  \bibfield  {author} {\bibinfo {author} {\bibfnamefont {C.}~\bibnamefont
  {Lupien}}, \bibinfo {author} {\bibfnamefont {W.~A.}\ \bibnamefont
  {MacFarlane}}, \bibinfo {author} {\bibfnamefont {Cyril}\ \bibnamefont
  {Proust}}, \bibinfo {author} {\bibfnamefont {Louis}\ \bibnamefont
  {Taillefer}}, \bibinfo {author} {\bibfnamefont {Z.~Q.}\ \bibnamefont {Mao}},
  \ and\ \bibinfo {author} {\bibfnamefont {Y.}~\bibnamefont {Maeno}},\
  }\bibfield  {title} {\enquote {\bibinfo {title} {Ultrasound attenuation in
  {S}r$_2${R}u{O}$_4$: An angle-resolved study of the superconducting gap
  function},}\ }\href {\doibase 10.1103/PhysRevLett.86.5986} {\bibfield
  {journal} {\bibinfo  {journal} {Phys. Rev. Lett.}\ }\textbf {\bibinfo
  {volume} {86}},\ \bibinfo {pages} {5986--5989} (\bibinfo {year}
  {2001})}\BibitemShut {NoStop}%
\bibitem [{\citenamefont {{Ghosh}}\ \emph {et~al.}(2020)\citenamefont
  {{Ghosh}}, \citenamefont {{Shekhter}}, \citenamefont {{Jerzembeck}},
  \citenamefont {{Kikugawa}}, \citenamefont {{Sokolov}}, \citenamefont
  {{Brando}}, \citenamefont {{Mackenzie}}, \citenamefont {{Hicks}},\ and\
  \citenamefont {{Ramshaw}}}]{Ghosh:2020}%
  \BibitemOpen
  \bibfield  {author} {\bibinfo {author} {\bibfnamefont {Sayak}\ \bibnamefont
  {{Ghosh}}}, \bibinfo {author} {\bibfnamefont {Arkady}\ \bibnamefont
  {{Shekhter}}}, \bibinfo {author} {\bibfnamefont {F.}~\bibnamefont
  {{Jerzembeck}}}, \bibinfo {author} {\bibfnamefont {N.}~\bibnamefont
  {{Kikugawa}}}, \bibinfo {author} {\bibfnamefont {Dmitry~A.}\ \bibnamefont
  {{Sokolov}}}, \bibinfo {author} {\bibfnamefont {Manuel}\ \bibnamefont
  {{Brando}}}, \bibinfo {author} {\bibfnamefont {A.~P.}\ \bibnamefont
  {{Mackenzie}}}, \bibinfo {author} {\bibfnamefont {Clifford~W.}\ \bibnamefont
  {{Hicks}}}, \ and\ \bibinfo {author} {\bibfnamefont {B.~J.}\ \bibnamefont
  {{Ramshaw}}},\ }\bibfield  {title} {\enquote {\bibinfo {title}
  {{Thermodynamic Evidence for a Two-Component Superconducting Order Parameter
  in Sr$_2$RuO$_4$}},}\ }\href@noop {} {\bibfield  {journal} {\bibinfo
  {journal} {Nature Physics}\ } (\bibinfo {year} {2020})}\BibitemShut {NoStop}%
\bibitem [{\citenamefont {Kidwingira}\ \emph {et~al.}(2006)\citenamefont
  {Kidwingira}, \citenamefont {Strand}, \citenamefont {Van~Harlingen},\ and\
  \citenamefont {Maeno}}]{Kidwingira:2006}%
  \BibitemOpen
  \bibfield  {author} {\bibinfo {author} {\bibfnamefont {Francoise}\
  \bibnamefont {Kidwingira}}, \bibinfo {author} {\bibfnamefont {J.~D.}\
  \bibnamefont {Strand}}, \bibinfo {author} {\bibfnamefont {D.~J.}\
  \bibnamefont {Van~Harlingen}}, \ and\ \bibinfo {author} {\bibfnamefont
  {Yoshiteru}\ \bibnamefont {Maeno}},\ }\bibfield  {title} {\enquote {\bibinfo
  {title} {Dynamical superconducting order parameter domains in
  {S}r$_2${R}u{O}$_4$},}\ }\href {\doibase 10.1126/science.1133239} {\bibfield
  {journal} {\bibinfo  {journal} {Science}\ }\textbf {\bibinfo {volume}
  {314}},\ \bibinfo {pages} {1267--1271} (\bibinfo {year} {2006})}\BibitemShut
  {NoStop}%
\bibitem [{\citenamefont {Anwar}\ \emph {et~al.}(2013)\citenamefont {Anwar},
  \citenamefont {Nakamura}, \citenamefont {Yonezawa}, \citenamefont {Yakabe},
  \citenamefont {Ishiguro}, \citenamefont {Takayanagi},\ and\ \citenamefont
  {Maeno}}]{Nakamura:2013}%
  \BibitemOpen
  \bibfield  {author} {\bibinfo {author} {\bibfnamefont {M.~S.}\ \bibnamefont
  {Anwar}}, \bibinfo {author} {\bibfnamefont {Taketomo}\ \bibnamefont
  {Nakamura}}, \bibinfo {author} {\bibfnamefont {S.}~\bibnamefont {Yonezawa}},
  \bibinfo {author} {\bibfnamefont {M.}~\bibnamefont {Yakabe}}, \bibinfo
  {author} {\bibfnamefont {R.}~\bibnamefont {Ishiguro}}, \bibinfo {author}
  {\bibfnamefont {H.}~\bibnamefont {Takayanagi}}, \ and\ \bibinfo {author}
  {\bibfnamefont {Y.}~\bibnamefont {Maeno}},\ }\bibfield  {title} {\enquote
  {\bibinfo {title} {Anomalous switching in {N}b/{R}u/{S}r$_2${R}u{O}$_4$
  topological junctions by chiral domain wall motion},}\ }\href@noop {}
  {\bibfield  {journal} {\bibinfo  {journal} {Scientific Reports}\ }\textbf
  {\bibinfo {volume} {3}},\ \bibinfo {pages} {2480} (\bibinfo {year}
  {2013})}\BibitemShut {NoStop}%
\bibitem [{\citenamefont {{Jerzembeck}}\ \emph {et~al.}(2021)\citenamefont
  {{Jerzembeck}}, \citenamefont {{R{\o}ising}}, \citenamefont {{Steppke}},
  \citenamefont {{Rosner}}, \citenamefont {{Sokolov}}, \citenamefont
  {{Kikugawa}}, \citenamefont {{Scaffidi}}, \citenamefont {{Simon}},
  \citenamefont {{Mackenzie}},\ and\ \citenamefont
  {{Hicks}}}]{Jerzembeck:2021}%
  \BibitemOpen
  \bibfield  {author} {\bibinfo {author} {\bibfnamefont {Fabian}\ \bibnamefont
  {{Jerzembeck}}}, \bibinfo {author} {\bibfnamefont {Henrik~S.}\ \bibnamefont
  {{R{\o}ising}}}, \bibinfo {author} {\bibfnamefont {Alexander}\ \bibnamefont
  {{Steppke}}}, \bibinfo {author} {\bibfnamefont {Helge}\ \bibnamefont
  {{Rosner}}}, \bibinfo {author} {\bibfnamefont {Dmitry~A.}\ \bibnamefont
  {{Sokolov}}}, \bibinfo {author} {\bibfnamefont {Naoki}\ \bibnamefont
  {{Kikugawa}}}, \bibinfo {author} {\bibfnamefont {Thomas}\ \bibnamefont
  {{Scaffidi}}}, \bibinfo {author} {\bibfnamefont {Steven~H.}\ \bibnamefont
  {{Simon}}}, \bibinfo {author} {\bibfnamefont {Andrew~P.}\ \bibnamefont
  {{Mackenzie}}}, \ and\ \bibinfo {author} {\bibfnamefont {Clifford~W.}\
  \bibnamefont {{Hicks}}},\ }\bibfield  {title} {\enquote {\bibinfo {title}
  {{The Superconductivity of {S}r$_2${R}u{O}$_4$ Under c-Axis Uniaxial
  Stress}},}\ }\href@noop {} {\bibfield  {journal} {\bibinfo  {journal} {arXiv
  e-prints}\ ,\ \bibinfo {eid} {arXiv:2111.06228}} (\bibinfo {year}
  {2021})}\BibitemShut {NoStop}%
\bibitem [{\citenamefont {Hicks}\ \emph
  {et~al.}(2014{\natexlab{b}})\citenamefont {Hicks}, \citenamefont {Brodsky},
  \citenamefont {Yelland}, \citenamefont {Gibbs}, \citenamefont {Bruin},
  \citenamefont {Barber}, \citenamefont {Edkins}, \citenamefont {Nishimura},
  \citenamefont {Yonezawa}, \citenamefont {Maeno},\ and\ \citenamefont
  {Mackenzie}}]{Hicks:2014110}%
  \BibitemOpen
  \bibfield  {author} {\bibinfo {author} {\bibfnamefont {Clifford~W.}\
  \bibnamefont {Hicks}}, \bibinfo {author} {\bibfnamefont {Daniel~O.}\
  \bibnamefont {Brodsky}}, \bibinfo {author} {\bibfnamefont {Edward~A.}\
  \bibnamefont {Yelland}}, \bibinfo {author} {\bibfnamefont {Alexandra~S.}\
  \bibnamefont {Gibbs}}, \bibinfo {author} {\bibfnamefont {Jan A.~N.}\
  \bibnamefont {Bruin}}, \bibinfo {author} {\bibfnamefont {Mark~E.}\
  \bibnamefont {Barber}}, \bibinfo {author} {\bibfnamefont {Stephen~D.}\
  \bibnamefont {Edkins}}, \bibinfo {author} {\bibfnamefont {Keigo}\
  \bibnamefont {Nishimura}}, \bibinfo {author} {\bibfnamefont {Shingo}\
  \bibnamefont {Yonezawa}}, \bibinfo {author} {\bibfnamefont {Yoshiteru}\
  \bibnamefont {Maeno}}, \ and\ \bibinfo {author} {\bibfnamefont {Andrew~P.}\
  \bibnamefont {Mackenzie}},\ }\bibfield  {title} {\enquote {\bibinfo {title}
  {Strong increase of <i>t</i><sub>c</sub> of {S}r$_2${R}u{O}$_4$ under both
  tensile and compressive strain},}\ }\href {\doibase 10.1126/science.1248292}
  {\bibfield  {journal} {\bibinfo  {journal} {Science}\ }\textbf {\bibinfo
  {volume} {344}},\ \bibinfo {pages} {283--285} (\bibinfo {year}
  {2014}{\natexlab{b}})}\BibitemShut {NoStop}%
\bibitem [{\citenamefont {Kittaka}\ \emph {et~al.}(2018)\citenamefont
  {Kittaka}, \citenamefont {Nakamura}, \citenamefont {Sakakibara},
  \citenamefont {Kikugawa}, \citenamefont {Terashima}, \citenamefont {Uji},
  \citenamefont {Sokolov}, \citenamefont {Mackenzie}, \citenamefont {Irie},
  \citenamefont {Tsutsumi}, \citenamefont {Suzuki},\ and\ \citenamefont
  {Machida}}]{Kittaka:2018}%
  \BibitemOpen
  \bibfield  {author} {\bibinfo {author} {\bibfnamefont {Shunichiro}\
  \bibnamefont {Kittaka}}, \bibinfo {author} {\bibfnamefont {Shota}\
  \bibnamefont {Nakamura}}, \bibinfo {author} {\bibfnamefont {Toshiro}\
  \bibnamefont {Sakakibara}}, \bibinfo {author} {\bibfnamefont {Naoki}\
  \bibnamefont {Kikugawa}}, \bibinfo {author} {\bibfnamefont {Taichi}\
  \bibnamefont {Terashima}}, \bibinfo {author} {\bibfnamefont {Shinya}\
  \bibnamefont {Uji}}, \bibinfo {author} {\bibfnamefont {Dmitry~A.}\
  \bibnamefont {Sokolov}}, \bibinfo {author} {\bibfnamefont {Andrew~P.}\
  \bibnamefont {Mackenzie}}, \bibinfo {author} {\bibfnamefont {Koki}\
  \bibnamefont {Irie}}, \bibinfo {author} {\bibfnamefont {Yasumasa}\
  \bibnamefont {Tsutsumi}}, \bibinfo {author} {\bibfnamefont {Katsuhiro}\
  \bibnamefont {Suzuki}}, \ and\ \bibinfo {author} {\bibfnamefont {Kazushige}\
  \bibnamefont {Machida}},\ }\bibfield  {title} {\enquote {\bibinfo {title}
  {Searching for gap zeros in {S}r$_2${R}u{O}$_4$ via field-angle-dependent
  specific-heat measurement},}\ }\href {\doibase 10.7566/JPSJ.87.093703}
  {\bibfield  {journal} {\bibinfo  {journal} {Journal of the Physical Society
  of Japan}\ }\textbf {\bibinfo {volume} {87}},\ \bibinfo {pages} {093703}
  (\bibinfo {year} {2018})}\BibitemShut {NoStop}%
\bibitem [{\citenamefont {Hassinger}\ \emph {et~al.}(2017)\citenamefont
  {Hassinger}, \citenamefont {Bourgeois-Hope}, \citenamefont {Taniguchi},
  \citenamefont {Ren\'e~de Cotret}, \citenamefont {Grissonnanche},
  \citenamefont {Anwar}, \citenamefont {Maeno}, \citenamefont
  {Doiron-Leyraud},\ and\ \citenamefont {Taillefer}}]{Hassinger:2017}%
  \BibitemOpen
  \bibfield  {author} {\bibinfo {author} {\bibfnamefont {E.}~\bibnamefont
  {Hassinger}}, \bibinfo {author} {\bibfnamefont {P.}~\bibnamefont
  {Bourgeois-Hope}}, \bibinfo {author} {\bibfnamefont {H.}~\bibnamefont
  {Taniguchi}}, \bibinfo {author} {\bibfnamefont {S.}~\bibnamefont {Ren\'e~de
  Cotret}}, \bibinfo {author} {\bibfnamefont {G.}~\bibnamefont
  {Grissonnanche}}, \bibinfo {author} {\bibfnamefont {M.~S.}\ \bibnamefont
  {Anwar}}, \bibinfo {author} {\bibfnamefont {Y.}~\bibnamefont {Maeno}},
  \bibinfo {author} {\bibfnamefont {N.}~\bibnamefont {Doiron-Leyraud}}, \ and\
  \bibinfo {author} {\bibfnamefont {Louis}\ \bibnamefont {Taillefer}},\
  }\bibfield  {title} {\enquote {\bibinfo {title} {Vertical line nodes in the
  superconducting gap structure of {S}r$_2${R}u{O}$_4$},}\ }\href {\doibase
  10.1103/PhysRevX.7.011032} {\bibfield  {journal} {\bibinfo  {journal} {Phys.
  Rev. X}\ }\textbf {\bibinfo {volume} {7}},\ \bibinfo {pages} {011032}
  (\bibinfo {year} {2017})}\BibitemShut {NoStop}%
\bibitem [{\citenamefont {Suh}\ \emph {et~al.}(2020)\citenamefont {Suh},
  \citenamefont {Menke}, \citenamefont {Brydon}, \citenamefont {Timm},
  \citenamefont {Ramires},\ and\ \citenamefont {Agterberg}}]{Suh:2020}%
  \BibitemOpen
  \bibfield  {author} {\bibinfo {author} {\bibfnamefont {Han~Gyeol}\
  \bibnamefont {Suh}}, \bibinfo {author} {\bibfnamefont {Henri}\ \bibnamefont
  {Menke}}, \bibinfo {author} {\bibfnamefont {P.~M.~R.}\ \bibnamefont
  {Brydon}}, \bibinfo {author} {\bibfnamefont {Carsten}\ \bibnamefont {Timm}},
  \bibinfo {author} {\bibfnamefont {Aline}\ \bibnamefont {Ramires}}, \ and\
  \bibinfo {author} {\bibfnamefont {Daniel~F.}\ \bibnamefont {Agterberg}},\
  }\bibfield  {title} {\enquote {\bibinfo {title} {Stabilizing even-parity
  chiral superconductivity in {S}r$_2${R}u{O}$_4$},}\ }\href {\doibase
  10.1103/PhysRevResearch.2.032023} {\bibfield  {journal} {\bibinfo  {journal}
  {Phys. Rev. Research}\ }\textbf {\bibinfo {volume} {2}},\ \bibinfo {pages}
  {032023} (\bibinfo {year} {2020})}\BibitemShut {NoStop}%
\bibitem [{\citenamefont {Ramires}\ and\ \citenamefont
  {Sigrist}(2016)}]{Ramires:2016}%
  \BibitemOpen
  \bibfield  {author} {\bibinfo {author} {\bibfnamefont {Aline}\ \bibnamefont
  {Ramires}}\ and\ \bibinfo {author} {\bibfnamefont {Manfred}\ \bibnamefont
  {Sigrist}},\ }\bibfield  {title} {\enquote {\bibinfo {title} {Identifying
  detrimental effects for multiorbital superconductivity: Application to
  {S}r$_2${R}u{O}$_4$},}\ }\href {\doibase 10.1103/PhysRevB.94.104501}
  {\bibfield  {journal} {\bibinfo  {journal} {Phys. Rev. B}\ }\textbf {\bibinfo
  {volume} {94}},\ \bibinfo {pages} {104501} (\bibinfo {year}
  {2016})}\BibitemShut {NoStop}%
\bibitem [{\citenamefont {Ramires}\ and\ \citenamefont
  {Sigrist}(2017)}]{Ramires:2017}%
  \BibitemOpen
  \bibfield  {author} {\bibinfo {author} {\bibfnamefont {Aline}\ \bibnamefont
  {Ramires}}\ and\ \bibinfo {author} {\bibfnamefont {Manfred}\ \bibnamefont
  {Sigrist}},\ }\bibfield  {title} {\enquote {\bibinfo {title} {A note on the
  upper critical field of {S}r$_2${R}u{O}$_4$ under strain},}\ }\href@noop {}
  {\bibfield  {journal} {\bibinfo  {journal} {J. Phys.: Conf. Ser.}\ }\textbf
  {\bibinfo {volume} {807}} (\bibinfo {year} {2017})}\BibitemShut {NoStop}%
\bibitem [{\citenamefont {Ramires}\ \emph {et~al.}(2018)\citenamefont
  {Ramires}, \citenamefont {Agterberg},\ and\ \citenamefont
  {Sigrist}}]{Ramires:2018}%
  \BibitemOpen
  \bibfield  {author} {\bibinfo {author} {\bibfnamefont {Aline}\ \bibnamefont
  {Ramires}}, \bibinfo {author} {\bibfnamefont {Daniel~F.}\ \bibnamefont
  {Agterberg}}, \ and\ \bibinfo {author} {\bibfnamefont {Manfred}\ \bibnamefont
  {Sigrist}},\ }\bibfield  {title} {\enquote {\bibinfo {title} {Tailoring
  ${T}_{c}$ by symmetry principles: The concept of superconducting fitness},}\
  }\href {\doibase 10.1103/PhysRevB.98.024501} {\bibfield  {journal} {\bibinfo
  {journal} {Phys. Rev. B}\ }\textbf {\bibinfo {volume} {98}},\ \bibinfo
  {pages} {024501} (\bibinfo {year} {2018})}\BibitemShut {NoStop}%
\bibitem [{\citenamefont {{Ramires}}(2021)}]{Ramires:2021}%
  \BibitemOpen
  \bibfield  {author} {\bibinfo {author} {\bibfnamefont {Aline}\ \bibnamefont
  {{Ramires}}},\ }\bibfield  {title} {\enquote {\bibinfo {title} {{Nodal gaps
  from local interactions in {S}r$_2${R}u{O}$_4$}},}\ }\href@noop {} {\bibfield
   {journal} {\bibinfo  {journal} {arXiv e-prints}\ ,\ \bibinfo {eid}
  {arXiv:2110.10621}} (\bibinfo {year} {2021})},\ \Eprint
  {http://arxiv.org/abs/2110.10621} {arXiv:2110.10621 [cond-mat.supr-con]}
  \BibitemShut {NoStop}%
\bibitem [{SM()}]{SM}%
  \BibitemOpen
  \href@noop {} {\bibinfo  {journal} {See Supplemental Material}\ }\BibitemShut
  {NoStop}%
\bibitem [{\citenamefont {Ramires}\ and\ \citenamefont
  {Sigrist}(2019)}]{Ramires:2019}%
  \BibitemOpen
\bibfield  {journal} {  }\bibfield  {author} {\bibinfo {author} {\bibfnamefont
  {Aline}\ \bibnamefont {Ramires}}\ and\ \bibinfo {author} {\bibfnamefont
  {Manfred}\ \bibnamefont {Sigrist}},\ }\bibfield  {title} {\enquote {\bibinfo
  {title} {Superconducting order parameter of {S}r$_2${R}u{O}$_4$: A
  microscopic perspective},}\ }\href {\doibase 10.1103/PhysRevB.100.104501}
  {\bibfield  {journal} {\bibinfo  {journal} {Phys. Rev. B}\ }\textbf {\bibinfo
  {volume} {100}},\ \bibinfo {pages} {104501} (\bibinfo {year}
  {2019})}\BibitemShut {NoStop}%
\bibitem [{\citenamefont {Vogt}\ and\ \citenamefont
  {Buttrey}(1995)}]{Vogt:1995}%
  \BibitemOpen
  \bibfield  {author} {\bibinfo {author} {\bibfnamefont {T.}~\bibnamefont
  {Vogt}}\ and\ \bibinfo {author} {\bibfnamefont {D.~J.}\ \bibnamefont
  {Buttrey}},\ }\bibfield  {title} {\enquote {\bibinfo {title} {Low-temperature
  structural behavior of {S}r$_2${R}u{O}$_4$},}\ }\href {\doibase
  10.1103/PhysRevB.52.R9843} {\bibfield  {journal} {\bibinfo  {journal} {Phys.
  Rev. B}\ }\textbf {\bibinfo {volume} {52}},\ \bibinfo {pages} {R9843--R9846}
  (\bibinfo {year} {1995})}\BibitemShut {NoStop}%
\bibitem [{\citenamefont {Barber}\ \emph {et~al.}(2019)\citenamefont {Barber},
  \citenamefont {Lechermann}, \citenamefont {Streltsov}, \citenamefont
  {Skornyakov}, \citenamefont {Ghosh}, \citenamefont {Ramshaw}, \citenamefont
  {Kikugawa}, \citenamefont {Sokolov}, \citenamefont {Mackenzie}, \citenamefont
  {Hicks},\ and\ \citenamefont {Mazin}}]{Barber:2019}%
  \BibitemOpen
  \bibfield  {author} {\bibinfo {author} {\bibfnamefont {Mark~E.}\ \bibnamefont
  {Barber}}, \bibinfo {author} {\bibfnamefont {Frank}\ \bibnamefont
  {Lechermann}}, \bibinfo {author} {\bibfnamefont {Sergey~V.}\ \bibnamefont
  {Streltsov}}, \bibinfo {author} {\bibfnamefont {Sergey~L.}\ \bibnamefont
  {Skornyakov}}, \bibinfo {author} {\bibfnamefont {Sayak}\ \bibnamefont
  {Ghosh}}, \bibinfo {author} {\bibfnamefont {B.~J.}\ \bibnamefont {Ramshaw}},
  \bibinfo {author} {\bibfnamefont {Naoki}\ \bibnamefont {Kikugawa}}, \bibinfo
  {author} {\bibfnamefont {Dmitry~A.}\ \bibnamefont {Sokolov}}, \bibinfo
  {author} {\bibfnamefont {Andrew~P.}\ \bibnamefont {Mackenzie}}, \bibinfo
  {author} {\bibfnamefont {Clifford~W.}\ \bibnamefont {Hicks}}, \ and\ \bibinfo
  {author} {\bibfnamefont {I.~I.}\ \bibnamefont {Mazin}},\ }\bibfield  {title}
  {\enquote {\bibinfo {title} {Role of correlations in determining the van hove
  strain in {S}r$_2${R}u{O}$_4$},}\ }\href {\doibase
  10.1103/PhysRevB.100.245139} {\bibfield  {journal} {\bibinfo  {journal}
  {Phys. Rev. B}\ }\textbf {\bibinfo {volume} {100}},\ \bibinfo {pages}
  {245139} (\bibinfo {year} {2019})}\BibitemShut {NoStop}%
\bibitem [{\citenamefont {Blaha}\ \emph {et~al.}()\citenamefont {Blaha},
  \citenamefont {Schwarz}, \citenamefont {HMadsen}, \citenamefont {Kvasnicka},
  \citenamefont {Luitz}, \citenamefont {Laskowsk}, \citenamefont {Tran},
  \citenamefont {Marks},\ and\ \citenamefont {Marks}}]{Blaha:2019}%
  \BibitemOpen
  \bibfield  {author} {\bibinfo {author} {\bibfnamefont {Peter}\ \bibnamefont
  {Blaha}}, \bibinfo {author} {\bibfnamefont {Karlheinz}\ \bibnamefont
  {Schwarz}}, \bibinfo {author} {\bibfnamefont {Georg~K}\ \bibnamefont
  {HMadsen}}, \bibinfo {author} {\bibfnamefont {Dieter}\ \bibnamefont
  {Kvasnicka}}, \bibinfo {author} {\bibfnamefont {Joachim}\ \bibnamefont
  {Luitz}}, \bibinfo {author} {\bibfnamefont {Robert}\ \bibnamefont
  {Laskowsk}}, \bibinfo {author} {\bibfnamefont {Fabien}\ \bibnamefont {Tran}},
  \bibinfo {author} {\bibfnamefont {Laurence}\ \bibnamefont {Marks}}, \ and\
  \bibinfo {author} {\bibfnamefont {Laurence}\ \bibnamefont {Marks}},\
  }\href@noop {} {\emph {\bibinfo {title} {WIEN2k: An Augmented Plane Wave Plus
  Local Orbitals Program for Calculating Crystal Properties}}}\BibitemShut
  {NoStop}%
\bibitem [{\citenamefont {Perdew}\ \emph {et~al.}(1996)\citenamefont {Perdew},
  \citenamefont {Burke},\ and\ \citenamefont {Ernzerhof}}]{Perdew:1996}%
  \BibitemOpen
  \bibfield  {author} {\bibinfo {author} {\bibfnamefont {John~P.}\ \bibnamefont
  {Perdew}}, \bibinfo {author} {\bibfnamefont {Kieron}\ \bibnamefont {Burke}},
  \ and\ \bibinfo {author} {\bibfnamefont {Matthias}\ \bibnamefont
  {Ernzerhof}},\ }\bibfield  {title} {\enquote {\bibinfo {title} {Generalized
  gradient approximation made simple},}\ }\href {\doibase
  10.1103/PhysRevLett.77.3865} {\bibfield  {journal} {\bibinfo  {journal}
  {Phys. Rev. Lett.}\ }\textbf {\bibinfo {volume} {77}},\ \bibinfo {pages}
  {3865--3868} (\bibinfo {year} {1996})}\BibitemShut {NoStop}%
\bibitem [{\citenamefont {Marzari}\ and\ \citenamefont
  {Vanderbilt}(1997)}]{Marzar:1997}%
  \BibitemOpen
  \bibfield  {author} {\bibinfo {author} {\bibfnamefont {Nicola}\ \bibnamefont
  {Marzari}}\ and\ \bibinfo {author} {\bibfnamefont {David}\ \bibnamefont
  {Vanderbilt}},\ }\bibfield  {title} {\enquote {\bibinfo {title} {Maximally
  localized generalized wannier functions for composite energy bands},}\ }\href
  {\doibase 10.1103/PhysRevB.56.12847} {\bibfield  {journal} {\bibinfo
  {journal} {Phys. Rev. B}\ }\textbf {\bibinfo {volume} {56}},\ \bibinfo
  {pages} {12847--12865} (\bibinfo {year} {1997})}\BibitemShut {NoStop}%
\bibitem [{\citenamefont {Souza}\ \emph {et~al.}(2001)\citenamefont {Souza},
  \citenamefont {Marzari},\ and\ \citenamefont {Vanderbilt}}]{Souza:2001}%
  \BibitemOpen
  \bibfield  {author} {\bibinfo {author} {\bibfnamefont {Ivo}\ \bibnamefont
  {Souza}}, \bibinfo {author} {\bibfnamefont {Nicola}\ \bibnamefont {Marzari}},
  \ and\ \bibinfo {author} {\bibfnamefont {David}\ \bibnamefont {Vanderbilt}},\
  }\bibfield  {title} {\enquote {\bibinfo {title} {Maximally localized wannier
  functions for entangled energy bands},}\ }\href {\doibase
  10.1103/PhysRevB.65.035109} {\bibfield  {journal} {\bibinfo  {journal} {Phys.
  Rev. B}\ }\textbf {\bibinfo {volume} {65}},\ \bibinfo {pages} {035109}
  (\bibinfo {year} {2001})}\BibitemShut {NoStop}%
\bibitem [{\citenamefont {Kuneš}\ \emph {et~al.}(2010)\citenamefont {Kuneš},
  \citenamefont {Arita}, \citenamefont {Wissgott}, \citenamefont {Toschi},
  \citenamefont {Ikeda},\ and\ \citenamefont {Held}}]{Kunes:2010}%
  \BibitemOpen
  \bibfield  {author} {\bibinfo {author} {\bibfnamefont {Jan}\ \bibnamefont
  {Kuneš}}, \bibinfo {author} {\bibfnamefont {Ryotaro}\ \bibnamefont {Arita}},
  \bibinfo {author} {\bibfnamefont {Philipp}\ \bibnamefont {Wissgott}},
  \bibinfo {author} {\bibfnamefont {Alessandro}\ \bibnamefont {Toschi}},
  \bibinfo {author} {\bibfnamefont {Hiroaki}\ \bibnamefont {Ikeda}}, \ and\
  \bibinfo {author} {\bibfnamefont {Karsten}\ \bibnamefont {Held}},\ }\bibfield
   {title} {\enquote {\bibinfo {title} {Wien2wannier: From linearized augmented
  plane waves to maximally localized wannier functions},}\ }\href {\doibase
  https://doi.org/10.1016/j.cpc.2010.08.005} {\bibfield  {journal} {\bibinfo
  {journal} {Computer Physics Communications}\ }\textbf {\bibinfo {volume}
  {181}},\ \bibinfo {pages} {1888 -- 1895} (\bibinfo {year}
  {2010})}\BibitemShut {NoStop}%
\bibitem [{\citenamefont {Pizzi}\ \emph {et~al.}(2020)\citenamefont {Pizzi},
  \citenamefont {Vitale}, \citenamefont {Arita}, \citenamefont {Blügel},
  \citenamefont {Freimuth}, \citenamefont {G{\'{e}}ranton}, \citenamefont
  {Gibertini}, \citenamefont {Gresch}, \citenamefont {Johnson}, \citenamefont
  {Koretsune}, \citenamefont {Iba{\~{n}}ez-Azpiroz}, \citenamefont {Lee},
  \citenamefont {Lihm}, \citenamefont {Marchand}, \citenamefont {Marrazzo},
  \citenamefont {Mokrousov}, \citenamefont {Mustafa}, \citenamefont {Nohara},
  \citenamefont {Nomura}, \citenamefont {Paulatto}, \citenamefont
  {Ponc{\'{e}}}, \citenamefont {Ponweiser}, \citenamefont {Qiao}, \citenamefont
  {Thöle}, \citenamefont {Tsirkin}, \citenamefont {Wierzbowska}, \citenamefont
  {Marzari}, \citenamefont {Vanderbilt}, \citenamefont {Souza}, \citenamefont
  {Mostofi},\ and\ \citenamefont {Yates}}]{Pizzi:2020}%
  \BibitemOpen
  \bibfield  {author} {\bibinfo {author} {\bibfnamefont {Giovanni}\
  \bibnamefont {Pizzi}}, \bibinfo {author} {\bibfnamefont {Valerio}\
  \bibnamefont {Vitale}}, \bibinfo {author} {\bibfnamefont {Ryotaro}\
  \bibnamefont {Arita}}, \bibinfo {author} {\bibfnamefont {Stefan}\
  \bibnamefont {Blügel}}, \bibinfo {author} {\bibfnamefont {Frank}\
  \bibnamefont {Freimuth}}, \bibinfo {author} {\bibfnamefont {Guillaume}\
  \bibnamefont {G{\'{e}}ranton}}, \bibinfo {author} {\bibfnamefont {Marco}\
  \bibnamefont {Gibertini}}, \bibinfo {author} {\bibfnamefont {Dominik}\
  \bibnamefont {Gresch}}, \bibinfo {author} {\bibfnamefont {Charles}\
  \bibnamefont {Johnson}}, \bibinfo {author} {\bibfnamefont {Takashi}\
  \bibnamefont {Koretsune}}, \bibinfo {author} {\bibfnamefont {Julen}\
  \bibnamefont {Iba{\~{n}}ez-Azpiroz}}, \bibinfo {author} {\bibfnamefont
  {Hyungjun}\ \bibnamefont {Lee}}, \bibinfo {author} {\bibfnamefont {Jae-Mo}\
  \bibnamefont {Lihm}}, \bibinfo {author} {\bibfnamefont {Daniel}\ \bibnamefont
  {Marchand}}, \bibinfo {author} {\bibfnamefont {Antimo}\ \bibnamefont
  {Marrazzo}}, \bibinfo {author} {\bibfnamefont {Yuriy}\ \bibnamefont
  {Mokrousov}}, \bibinfo {author} {\bibfnamefont {Jamal~I}\ \bibnamefont
  {Mustafa}}, \bibinfo {author} {\bibfnamefont {Yoshiro}\ \bibnamefont
  {Nohara}}, \bibinfo {author} {\bibfnamefont {Yusuke}\ \bibnamefont {Nomura}},
  \bibinfo {author} {\bibfnamefont {Lorenzo}\ \bibnamefont {Paulatto}},
  \bibinfo {author} {\bibfnamefont {Samuel}\ \bibnamefont {Ponc{\'{e}}}},
  \bibinfo {author} {\bibfnamefont {Thomas}\ \bibnamefont {Ponweiser}},
  \bibinfo {author} {\bibfnamefont {Junfeng}\ \bibnamefont {Qiao}}, \bibinfo
  {author} {\bibfnamefont {Florian}\ \bibnamefont {Thöle}}, \bibinfo {author}
  {\bibfnamefont {Stepan~S}\ \bibnamefont {Tsirkin}}, \bibinfo {author}
  {\bibfnamefont {Ma{\l}gorzata}\ \bibnamefont {Wierzbowska}}, \bibinfo
  {author} {\bibfnamefont {Nicola}\ \bibnamefont {Marzari}}, \bibinfo {author}
  {\bibfnamefont {David}\ \bibnamefont {Vanderbilt}}, \bibinfo {author}
  {\bibfnamefont {Ivo}\ \bibnamefont {Souza}}, \bibinfo {author} {\bibfnamefont
  {Arash~A}\ \bibnamefont {Mostofi}}, \ and\ \bibinfo {author} {\bibfnamefont
  {Jonathan~R}\ \bibnamefont {Yates}},\ }\bibfield  {title} {\enquote {\bibinfo
  {title} {Wannier90 as a community code: new features and applications},}\
  }\href {\doibase 10.1088/1361-648x/ab51ff} {\bibfield  {journal} {\bibinfo
  {journal} {Journal of Physics: Condensed Matter}\ }\textbf {\bibinfo {volume}
  {32}},\ \bibinfo {pages} {165902} (\bibinfo {year} {2020})}\BibitemShut
  {NoStop}%
\bibitem [{\citenamefont {Parcollet}\ \emph {et~al.}(2015)\citenamefont
  {Parcollet}, \citenamefont {Ferrero}, \citenamefont {Ayral}, \citenamefont
  {Hafermann}, \citenamefont {Krivenko}, \citenamefont {Messio},\ and\
  \citenamefont {Seth}}]{Parcollet:2015}%
  \BibitemOpen
  \bibfield  {author} {\bibinfo {author} {\bibfnamefont {Olivier}\ \bibnamefont
  {Parcollet}}, \bibinfo {author} {\bibfnamefont {Michel}\ \bibnamefont
  {Ferrero}}, \bibinfo {author} {\bibfnamefont {Thomas}\ \bibnamefont {Ayral}},
  \bibinfo {author} {\bibfnamefont {Hartmut}\ \bibnamefont {Hafermann}},
  \bibinfo {author} {\bibfnamefont {Igor}\ \bibnamefont {Krivenko}}, \bibinfo
  {author} {\bibfnamefont {Laura}\ \bibnamefont {Messio}}, \ and\ \bibinfo
  {author} {\bibfnamefont {Priyanka}\ \bibnamefont {Seth}},\ }\bibfield
  {title} {\enquote {\bibinfo {title} {Triqs: A toolbox for research on
  interacting quantum systems},}\ }\href {\doibase
  https://doi.org/10.1016/j.cpc.2015.04.023} {\bibfield  {journal} {\bibinfo
  {journal} {Computer Physics Communications}\ }\textbf {\bibinfo {volume}
  {196}},\ \bibinfo {pages} {398 -- 415} (\bibinfo {year} {2015})}\BibitemShut
  {NoStop}%
\bibitem [{\citenamefont {Giannozzi}\ \emph {et~al.}(2009)\citenamefont
  {Giannozzi}, \citenamefont {Baroni}, \citenamefont {Bonini}, \citenamefont
  {Calandra}, \citenamefont {Car}, \citenamefont {Cavazzoni}, \citenamefont
  {Ceresoli}, \citenamefont {Chiarotti}, \citenamefont {Cococcioni},
  \citenamefont {Dabo}, \citenamefont {Corso}, \citenamefont {de~Gironcoli},
  \citenamefont {Fabris}, \citenamefont {Fratesi}, \citenamefont {Gebauer},
  \citenamefont {Gerstmann}, \citenamefont {Gougoussis}, \citenamefont
  {Kokalj}, \citenamefont {Lazzeri}, \citenamefont {Martin-Samos},
  \citenamefont {Marzari}, \citenamefont {Mauri}, \citenamefont {Mazzarello},
  \citenamefont {Paolini}, \citenamefont {Pasquarello}, \citenamefont
  {Paulatto}, \citenamefont {Sbraccia}, \citenamefont {Scandolo}, \citenamefont
  {Sclauzero}, \citenamefont {Seitsonen}, \citenamefont {Smogunov},
  \citenamefont {Umari},\ and\ \citenamefont {Wentzcovitch}}]{Giannozzi:2009}%
  \BibitemOpen
  \bibfield  {author} {\bibinfo {author} {\bibfnamefont {Paolo}\ \bibnamefont
  {Giannozzi}}, \bibinfo {author} {\bibfnamefont {Stefano}\ \bibnamefont
  {Baroni}}, \bibinfo {author} {\bibfnamefont {Nicola}\ \bibnamefont {Bonini}},
  \bibinfo {author} {\bibfnamefont {Matteo}\ \bibnamefont {Calandra}}, \bibinfo
  {author} {\bibfnamefont {Roberto}\ \bibnamefont {Car}}, \bibinfo {author}
  {\bibfnamefont {Carlo}\ \bibnamefont {Cavazzoni}}, \bibinfo {author}
  {\bibfnamefont {Davide}\ \bibnamefont {Ceresoli}}, \bibinfo {author}
  {\bibfnamefont {Guido~L}\ \bibnamefont {Chiarotti}}, \bibinfo {author}
  {\bibfnamefont {Matteo}\ \bibnamefont {Cococcioni}}, \bibinfo {author}
  {\bibfnamefont {Ismaila}\ \bibnamefont {Dabo}}, \bibinfo {author}
  {\bibfnamefont {Andrea~Dal}\ \bibnamefont {Corso}}, \bibinfo {author}
  {\bibfnamefont {Stefano}\ \bibnamefont {de~Gironcoli}}, \bibinfo {author}
  {\bibfnamefont {Stefano}\ \bibnamefont {Fabris}}, \bibinfo {author}
  {\bibfnamefont {Guido}\ \bibnamefont {Fratesi}}, \bibinfo {author}
  {\bibfnamefont {Ralph}\ \bibnamefont {Gebauer}}, \bibinfo {author}
  {\bibfnamefont {Uwe}\ \bibnamefont {Gerstmann}}, \bibinfo {author}
  {\bibfnamefont {Christos}\ \bibnamefont {Gougoussis}}, \bibinfo {author}
  {\bibfnamefont {Anton}\ \bibnamefont {Kokalj}}, \bibinfo {author}
  {\bibfnamefont {Michele}\ \bibnamefont {Lazzeri}}, \bibinfo {author}
  {\bibfnamefont {Layla}\ \bibnamefont {Martin-Samos}}, \bibinfo {author}
  {\bibfnamefont {Nicola}\ \bibnamefont {Marzari}}, \bibinfo {author}
  {\bibfnamefont {Francesco}\ \bibnamefont {Mauri}}, \bibinfo {author}
  {\bibfnamefont {Riccardo}\ \bibnamefont {Mazzarello}}, \bibinfo {author}
  {\bibfnamefont {Stefano}\ \bibnamefont {Paolini}}, \bibinfo {author}
  {\bibfnamefont {Alfredo}\ \bibnamefont {Pasquarello}}, \bibinfo {author}
  {\bibfnamefont {Lorenzo}\ \bibnamefont {Paulatto}}, \bibinfo {author}
  {\bibfnamefont {Carlo}\ \bibnamefont {Sbraccia}}, \bibinfo {author}
  {\bibfnamefont {Sandro}\ \bibnamefont {Scandolo}}, \bibinfo {author}
  {\bibfnamefont {Gabriele}\ \bibnamefont {Sclauzero}}, \bibinfo {author}
  {\bibfnamefont {Ari~P}\ \bibnamefont {Seitsonen}}, \bibinfo {author}
  {\bibfnamefont {Alexander}\ \bibnamefont {Smogunov}}, \bibinfo {author}
  {\bibfnamefont {Paolo}\ \bibnamefont {Umari}}, \ and\ \bibinfo {author}
  {\bibfnamefont {Renata~M}\ \bibnamefont {Wentzcovitch}},\ }\bibfield  {title}
  {\enquote {\bibinfo {title} {{QUANTUM} {ESPRESSO}: a modular and open-source
  software project for quantum simulations of materials},}\ }\href {\doibase
  10.1088/0953-8984/21/39/395502} {\bibfield  {journal} {\bibinfo  {journal}
  {Journal of Physics: Condensed Matter}\ }\textbf {\bibinfo {volume} {21}},\
  \bibinfo {pages} {395502} (\bibinfo {year} {2009})}\BibitemShut {NoStop}%
\bibitem [{\citenamefont {Garrity}\ \emph {et~al.}(2014)\citenamefont
  {Garrity}, \citenamefont {Bennett}, \citenamefont {Rabe},\ and\ \citenamefont
  {Vanderbilt}}]{Garrity:2014}%
  \BibitemOpen
  \bibfield  {author} {\bibinfo {author} {\bibfnamefont {Kevin~F.}\
  \bibnamefont {Garrity}}, \bibinfo {author} {\bibfnamefont {Joseph~W.}\
  \bibnamefont {Bennett}}, \bibinfo {author} {\bibfnamefont {Karin~M.}\
  \bibnamefont {Rabe}}, \ and\ \bibinfo {author} {\bibfnamefont {David}\
  \bibnamefont {Vanderbilt}},\ }\bibfield  {title} {\enquote {\bibinfo {title}
  {Pseudopotentials for high-throughput dft calculations},}\ }\href {\doibase
  https://doi.org/10.1016/j.commatsci.2013.08.053} {\bibfield  {journal}
  {\bibinfo  {journal} {Computational Materials Science}\ }\textbf {\bibinfo
  {volume} {81}},\ \bibinfo {pages} {446 -- 452} (\bibinfo {year}
  {2014})}\BibitemShut {NoStop}%
\bibitem [{\citenamefont {Min}\ \emph {et~al.}(2006)\citenamefont {Min},
  \citenamefont {Hill}, \citenamefont {Sinitsyn}, \citenamefont {Sahu},
  \citenamefont {Kleinman},\ and\ \citenamefont {MacDonald}}]{Min:2006}%
  \BibitemOpen
  \bibfield  {author} {\bibinfo {author} {\bibfnamefont {Hongki}\ \bibnamefont
  {Min}}, \bibinfo {author} {\bibfnamefont {J.~E.}\ \bibnamefont {Hill}},
  \bibinfo {author} {\bibfnamefont {N.~A.}\ \bibnamefont {Sinitsyn}}, \bibinfo
  {author} {\bibfnamefont {B.~R.}\ \bibnamefont {Sahu}}, \bibinfo {author}
  {\bibfnamefont {Leonard}\ \bibnamefont {Kleinman}}, \ and\ \bibinfo {author}
  {\bibfnamefont {A.~H.}\ \bibnamefont {MacDonald}},\ }\bibfield  {title}
  {\enquote {\bibinfo {title} {Intrinsic and rashba spin-orbit interactions in
  graphene sheets},}\ }\href {\doibase 10.1103/PhysRevB.74.165310} {\bibfield
  {journal} {\bibinfo  {journal} {Phys. Rev. B}\ }\textbf {\bibinfo {volume}
  {74}},\ \bibinfo {pages} {165310} (\bibinfo {year} {2006})}\BibitemShut
  {NoStop}%
\bibitem [{\citenamefont {Sigrist}\ and\ \citenamefont
  {Ueda}(1991)}]{Sigrist:1991}%
  \BibitemOpen
  \bibfield  {author} {\bibinfo {author} {\bibfnamefont {Manfred}\ \bibnamefont
  {Sigrist}}\ and\ \bibinfo {author} {\bibfnamefont {Kazuo}\ \bibnamefont
  {Ueda}},\ }\bibfield  {title} {\enquote {\bibinfo {title} {Phenomenological
  theory of unconventional superconductivity},}\ }\href {\doibase
  10.1103/RevModPhys.63.239} {\bibfield  {journal} {\bibinfo  {journal} {Rev.
  Mod. Phys.}\ }\textbf {\bibinfo {volume} {63}},\ \bibinfo {pages} {239--311}
  (\bibinfo {year} {1991})}\BibitemShut {NoStop}%
\bibitem [{\citenamefont {Tamai}\ \emph {et~al.}(2019)\citenamefont {Tamai},
  \citenamefont {Zingl}, \citenamefont {Rozbicki}, \citenamefont {Cappelli},
  \citenamefont {Ricc\`o}, \citenamefont {de~la Torre}, \citenamefont
  {McKeown~Walker}, \citenamefont {Bruno}, \citenamefont {King}, \citenamefont
  {Meevasana}, \citenamefont {Shi}, \citenamefont
  {Radovi\ifmmode~\acute{c}\else \'{c}\fi{}}, \citenamefont {Plumb},
  \citenamefont {Gibbs}, \citenamefont {Mackenzie}, \citenamefont {Berthod},
  \citenamefont {Strand}, \citenamefont {Kim}, \citenamefont {Georges},\ and\
  \citenamefont {Baumberger}}]{Tamai:2019}%
  \BibitemOpen
  \bibfield  {author} {\bibinfo {author} {\bibfnamefont {A.}~\bibnamefont
  {Tamai}}, \bibinfo {author} {\bibfnamefont {M.}~\bibnamefont {Zingl}},
  \bibinfo {author} {\bibfnamefont {E.}~\bibnamefont {Rozbicki}}, \bibinfo
  {author} {\bibfnamefont {E.}~\bibnamefont {Cappelli}}, \bibinfo {author}
  {\bibfnamefont {S.}~\bibnamefont {Ricc\`o}}, \bibinfo {author} {\bibfnamefont
  {A.}~\bibnamefont {de~la Torre}}, \bibinfo {author} {\bibfnamefont
  {S.}~\bibnamefont {McKeown~Walker}}, \bibinfo {author} {\bibfnamefont
  {F.~Y.}\ \bibnamefont {Bruno}}, \bibinfo {author} {\bibfnamefont {P.~D.~C.}\
  \bibnamefont {King}}, \bibinfo {author} {\bibfnamefont {W.}~\bibnamefont
  {Meevasana}}, \bibinfo {author} {\bibfnamefont {M.}~\bibnamefont {Shi}},
  \bibinfo {author} {\bibfnamefont {M.}~\bibnamefont
  {Radovi\ifmmode~\acute{c}\else \'{c}\fi{}}}, \bibinfo {author} {\bibfnamefont
  {N.~C.}\ \bibnamefont {Plumb}}, \bibinfo {author} {\bibfnamefont {A.~S.}\
  \bibnamefont {Gibbs}}, \bibinfo {author} {\bibfnamefont {A.~P.}\ \bibnamefont
  {Mackenzie}}, \bibinfo {author} {\bibfnamefont {C.}~\bibnamefont {Berthod}},
  \bibinfo {author} {\bibfnamefont {H.~U.~R.}\ \bibnamefont {Strand}}, \bibinfo
  {author} {\bibfnamefont {M.}~\bibnamefont {Kim}}, \bibinfo {author}
  {\bibfnamefont {A.}~\bibnamefont {Georges}}, \ and\ \bibinfo {author}
  {\bibfnamefont {F.}~\bibnamefont {Baumberger}},\ }\bibfield  {title}
  {\enquote {\bibinfo {title} {High-resolution photoemission on
  {S}r$_2${R}u{O}$_4$ reveals correlation-enhanced effective spin-orbit
  coupling and dominantly local self-energies},}\ }\href {\doibase
  10.1103/PhysRevX.9.021048} {\bibfield  {journal} {\bibinfo  {journal} {Phys.
  Rev. X}\ }\textbf {\bibinfo {volume} {9}},\ \bibinfo {pages} {021048}
  (\bibinfo {year} {2019})}\BibitemShut {NoStop}%
\bibitem [{\citenamefont {Klauss}()}]{Henning:2021}%
  \BibitemOpen
  \bibfield  {author} {\bibinfo {author} {\bibfnamefont {H.-H.}\ \bibnamefont
  {Klauss}},\ }\href@noop {} {\bibinfo  {journal} {private communication}\
  }\BibitemShut {NoStop}%
\end{thebibliography}%




\section*{Supplementary material}

# Supplemental Material - The effects of strain in multi-orbital superconductors: the case of $Sr_2RuO_4$

(Dated: November 26, 2021)

## COMPLETE NORMAL STATE HAMILTONIAN AND ORDER PARAMETERS FOR THE $t_{2g}$ MANIFOLD OF $Sr_2RuO_4$ IN THREE DIMENSIONS

Following the notation introduced in [1, 2], we describe the normal state of $Sr_2RuO_4$ by a three-dimensional three-orbital model. We define the spinor operator $\Phi_{\mathbf{k}}^{\dagger} = (c_{\mathbf{k},yz\uparrow}^{\dagger}, c_{\mathbf{k},yz\downarrow}^{\dagger}, c_{\mathbf{k},xz\uparrow}^{\dagger}, c_{\mathbf{k},xz\downarrow}^{\dagger}, c_{\mathbf{k},xy\uparrow}^{\dagger}, c_{\mathbf{k},xy\downarrow}^{\dagger})$, where $c_{\mathbf{k},\gamma\sigma}^{\dagger}$ creates an electron with momentum $\mathbf{k}$, orbital $\gamma$, and spin $\sigma$, in terms of which we can write the most general single-particle Hamiltonian

$$\mathcal{H}_0 = \sum_{\mathbf{k}} \Phi_{\mathbf{k}}^{\dagger} \hat{H}_0(\mathbf{k}) \Phi_{\mathbf{k}}, \tag{1}$$

with

$$\hat{H}_0(\mathbf{k}) = \sum_{a=0}^{8} \sum_{b=0}^{3} h_{ab}(\mathbf{k}) \, \lambda_a \otimes \sigma_b. \tag{2}$$

Here $\lambda_a$ are Gell-Mann matrices encoding the orbital degrees of freedom (DOF), $\sigma_b$ are Pauli matrices encoding the spin DOF ($\lambda_0$ and $\sigma_0$ are unit matrices), and $h_{ab}(\mathbf{k})$ are real and even functions of momentum. In presence of spatial inversion and time-reversal symmetries, out of the 36 pairs of indexes $(a, b)$, only 15 are symmetry allowed. These are enumerated in Table I, where we also highlight the associated physical process, and the irreducible representation (irrep) in $D_{4h}$ for the non-strained material and in $D_{2h}$ ($D_{2h}^*$) for in-plane uniaxial strain along the $\langle 100 \rangle$ ($\langle 110 \rangle$) direction.

In analogy to the normal state Hamiltonian, we parametrize the superconducting order parameters as

$$\hat{\Delta} = d_0 \, \lambda_a \otimes \sigma_b \, (i\sigma_2), \tag{3}$$

with magnitude $d_0$. Each order parameter component is labelled by a pair of indexes $[a, b]$ (note the different brackets used for the parametrization of the terms in the normal state Hamiltonian and for the order parameter). Here we focus on local, s-wave, order parameters enumerated in Table II, highlighting the irrep in $D_{4h}$, $D_{2h}$, and $D_{2h}^*$, and the orbital and spin character.

| Irrep in $D_{4h}$ | Irrep in $D_{2h}$ | Irrep in $D_{2h}*$ | $(a,b)$ | Physical process |
|---|---|---|---|---|
| $A_{1g}$ | $A_g$ | $A_g$ | $(0,0)$ | intra-orbital hopping |
| | | | $(8,0)$ | intra-orbital hopping |
| | | | $(4,3)$ | atomic SOC |
| | | | $(5,2)-(6,1)$ | atomic-SOC |
| $A_{2g}$ | $B_{1g}$ | $B_{1g}$ | $(5,1)+(6,2)$ | **k**-SOC |
| $B_{1g}$ | $A_g$ | $B_{1g}$ | $(7,0)$ | intra-orbital hopping |
| | | | $(5,2)+(6,1)$ | **k**-SOC |
| $B_{2g}$ | $B_{1g}$ | $A_g$ | $(1,0)$ | inter-orbital hopping |
| | | | $(5,1)-(6,2)$ | **k**-SOC |
| $E_g$ | $\{B_{3g},B_{2g}\}$ | $\{B_{3g},B_{2g}\}$ | $\{(3,0),-(2,0)\}$ | inter-orbital hopping |
| | | | $\{(4,2),-(4,1)\}$ | **k**-SOC |
| | | | $\{(5,3),(6,3)\}$ | **k**-SOC |

TABLE I. Symmetry-allowed $(a,b)$ terms in the normal-state Hamiltonian, the respective irrep in $D_{4h}$, $D_{2h}$ (for strain along $\langle 100 \rangle$), and $D_{2h}*$ (for strain along $\langle 110 \rangle$), and the associated physical process. For the two-dimensional irrep $E_g$ in $D_{4h}$, the entries are organized such that the first transforms as $x$ (or $yz$) and the second as $y$ (or $xz$). Once the group is reduced, the first entry transforms as $B_{3g}$ and the second as $B_{2g}$.

| Irrep in $D_{4h}$ | Irrep in $D_{2h}$ | Irrep in $D_{2h}*$ | $[a,b]$ | Orbital | Spin |
|---|---|---|---|---|---|
| $A_{1g}$ | $A_g$ | $A_g$ | $[0,0]$ | symmetric | singlet |
| | | | $[8,0]$ | symmetric | singlet |
| | | | $[4,3]$ | antisymmetric | triplet |
| | | | $[5,2]-[6,1]$ | antisymmetric | triplet |
| $A_{2g}$ | $B_{1g}$ | $B_{1g}$ | $[5,1]+[6,2]$ | antisymmetric | triplet |
| $B_{1g}$ | $A_g$ | $B_{1g}$ | $[7,0]$ | symmetric | singlet |
| | | | $[5,2]+[6,1]$ | antisymmetric | triplet |
| $B_{2g}$ | $B_{1g}$ | $A_{1g}$ | $[1,0]$ | symmetric | singlet |
| | | | $[5,1]-[6,2]$ | antisymmetric | triplet |
| $E_g$ | $\{B_{3g}, B_{2g}\}$ | $\{B_{3g}, B_{2g}\}$ | $\{[3,0],-[2,0]\}$ | symmetric | singlet |
| | | | $\{[4,2],-[4,1]\}$ | antisymmetric | triplet |
| | | | $\{[5,3],[6,3]\}$ | antisymmetric | triplet |

TABLE II. Local gap functions and irreps in $D_{4h}$, $D_{2h}$ (for strain along $\langle 100 \rangle$), and $D_{2h}*$ (for strain along $\langle 110 \rangle$). Highlighted are also the orbital and spin character.

**Strain along the $\langle 100 \rangle$ direction**

Uniaxial strain does not break inversion or time-reversal symmetry, therefore the analysis of the Hamiltonian and local gap structures in presence of strain falls back on the same set of 15 terms enumerated in Tables I and II [1, 2]. For strain along the $\langle 100 \rangle$ direction, the point group symmetry is reduced, such that each term is associated with a different irreducible representation. The reduction $D_{4h} \rightarrow D_{2h}$ leads to the mapping $A_{1g}/B_{1g} \rightarrow A_g$, $A_{2g}/B_{2g} \rightarrow B_{1g}$ and $E_g \rightarrow \{B_{2g}, B_{3g}\}$, as indicated in the second column of Tables I and II. This type of strain is usually referred to as $\epsilon_{xx} - \epsilon_{yy}$, with $B_{1g}$ symmetry, from which it is evident the non-equivalence between the $x$ and $y$ directions (an exchange of directions leads to an overall minus sign). In this case a splitting of the superconducting critical temperatures of a state associated with a two-dimensional irreducible representation is expected.

**Strain along the $\langle 001 \rangle$ direction**

For strain along the $\langle 001 \rangle$ direction there is no symmetry reduction. This type of strain is usually referred to as $\epsilon_{zz}$, with $A_{1g}$ symmetry, from which it is evident the equivalence between the $x$ and $y$ directions. In this case a splitting of the superconducting critical temperatures of a state associated with a two-dimensional irreducible representation is not expected.

**Strain along the $\langle 110 \rangle$ direction**

For strain along the $\langle 110 \rangle$ direction, the point group symmetry is reduced as $D_{4h} \rightarrow D_{2h}^*$, but now the remaining symmetry axes are different from the case of strain along the $\langle 100 \rangle$. This difference leads to a distinct mapping of irreps: $A_{1g}/B_{2g} \rightarrow A_g$, $A_{2g}/B_{1g} \rightarrow B_{1g}$ and $E_g \rightarrow \{B_{2g}, B_{3g}\}$, as indicated in the third column of Tables I and II. This type of strain is usually referred to as $\epsilon_{xy}$, with $B_{2g}$ symmetry, from which it is evident the equivalence between the $x$ and $y$ directions. In this case a splitting of the superconducting critical temperatures of a state associated with a two-dimensional irreducible representation is expected, but it cannot be captured by our simple models projected into the $XZ$ or $YZ$ planes, as discussed below.

## REDUCED TWO-ORBITAL MODELS ALONG THE $XZ$ AND $YZ$ PLANES

Before moving to the superconducting fitness analysis, here we introduce simplified two-orbital models. Note that, with the exception of the regions around the Brillouin zone diagonal, most paths stemming from the gamma point towards the Brillouin zone edge cut only two Fermi surfaces. For example, along the $XZ$ ($YZ$) plane we find only the $\beta$ and $\gamma$ bands, the former composed mainly of $d_{xz}$ ($d_{yz}$) and the later of $d_{xy}$ orbitals.

### Model along the $XZ$ plane

Along the $XZ$ plane, the dominant orbitals at the Fermi energy are $d_{xz}$ and $d_{xy}$. Projecting into this subspace, we obtain an effective two-orbital Hamiltonian:

$$\mathcal{H}_0^{XZ} = \sum_{\mathbf{k}} \Psi_{\mathbf{k}}^\dagger \hat{H}_0^{XZ}(\mathbf{k}) \Psi_{\mathbf{k}}, \tag{4}$$

in the basis $\Psi_{\mathbf{k}}^\dagger = (c_{\mathbf{k},xz\uparrow}^\dagger, c_{\mathbf{k},xz\downarrow}^\dagger, c_{\mathbf{k},xy\uparrow}^\dagger, c_{\mathbf{k},xy\downarrow}^\dagger)$, with

$$\hat{H}_0^{XZ}(\mathbf{k}) = \sum_{a,b=0}^{3} \tilde{h}_{ab}^{XZ}(\mathbf{k})\, \tau_a \otimes \sigma_b, \tag{5}$$

where the $\tilde{h}_{ab}^{XZ}(\mathbf{k})$ are real even functions of momentum, and both $\tau_a$ and $\sigma_b$ are Pauli matrices encoding the orbital and the spin DOF, respectively ($\sigma_0$ and $\tau_0$ are identity matrices). There are, in principle, 16 parameters $\tilde{h}_{ab}(\mathbf{k})$ but in the presence of time-reversal and inversion symmetries these are constrained to only six, including the term proportional to the identity. The symmetry-allowed terms are listed in Table III; we classify them in terms of the irreps of $D_{2h}$, which is the little group along the $XZ$ plane. We indicate the pair of indexes $(a, b)_{XZ}$ with a subindex to distinguish these from the original three-orbital model. The associated terms $(a, b)$ and irrep in $D_{4h}$ for the original three-orbital model are given, as well as the associated physical process.

The explicit form of the parameters $\tilde{h}_{ab}^{XZ}(\mathbf{k})$ is:

$$\tilde{h}_{00}^{XZ}(\mathbf{k}) = [\xi_{xz}(\mathbf{k}) + \xi_{xy}(\mathbf{k})]/2 \tag{6}$$

$$\tilde{h}_{10}^{XZ}(\mathbf{k}) = \xi_{xz/xy}(\mathbf{k}) = 0 \qquad \text{at the } k_y = 0 \text{ plane}$$

$$\tilde{h}_{21}^{XZ}(\mathbf{k}) = \eta$$

$$\tilde{h}_{22}^{XZ}(\mathbf{k}) = \eta_{xz/xy}^2(\mathbf{k}) = 0 \qquad \text{at the } k_y = 0 \text{ plane}$$

$$\tilde{h}_{23}^{XZ}(\mathbf{k}) = \eta_{xz/xy}^3(\mathbf{k})$$

$$\tilde{h}_{30}^{XZ}(\mathbf{k}) = [\xi_{xz}(\mathbf{k}) - \xi_{xy}(\mathbf{k})]/2,$$

with

$$\xi_{xz}(\mathbf{k}) = 2t_{xz}(\hat{x})\cos(k_x a) + 2t_{xz}(\hat{y})\cos k_y a + 2t_{xz}(\hat{d}_p)\cos(k_x a + k_y a) + 2t_{xz}(\hat{\bar{d}}_p)\cos(k_x a - k_y a)$$

$$\xi_{xy}(\mathbf{k}) = 2t_{xy}(\hat{x})\cos(k_x a) + 2t_{xy}(\hat{y})\cos k_y a + 2t_{xy}(\hat{d}_p)\cos(k_x a + k_y a) + 2t_{xy}(\hat{\bar{d}}_p)\cos(k_x a - k_y a)$$

$$\xi_{xz/xy}(\mathbf{k}) = -4t_{xz/xy}(\hat{d}_d)\sin\left(\frac{k_x + k_y}{2}\right)\sin\left(\frac{k_z}{2}\right) + 4t_{xz/xy}(\hat{\bar{d}}_d)\sin\left(\frac{k_x - k_y}{2}\right)\sin\left(\frac{k_z}{2}\right)$$

$$\eta_{xz/xy}^2(\mathbf{k}) = 2\lambda_{xz/xy}(\hat{d}_p)\cos(k_x + k_y) - 2\lambda_{xz/xy}(\hat{\bar{d}}_p)\cos(k_x - k_y)$$

$$\eta_{xz/xy}^3(\mathbf{k}) = -4\lambda_{xz/xy}(\hat{d}_d)\sin\left(\frac{k_x + k_y}{2}\right)\sin\left(\frac{k_z}{2}\right) - 4\lambda_{xz/xy}(\hat{\bar{d}}_d)\sin\left(\frac{k_x - k_y}{2}\right)\sin\left(\frac{k_z}{2}\right).$$

Here $t_\gamma(\hat{n})$ corresponds to the intra-orbital hopping amplitude for orbital $\gamma$ and $t_{\gamma/\gamma'}(\hat{n})$ corresponds to inter-orbital hopping between orbitals $\gamma$ and $\gamma'$ in direction $\hat{n}$. The directions correspond to $\hat{x} = (100)$, $\hat{y} = (010)$, $\hat{d}_p = (110)$, $\hat{\bar{d}}_p = (1-10)$, $\hat{d}_d = (111)$, $\hat{\bar{d}}_d = (1-11)$. Under strain $t \to t + Rs$, where $R$ is the rate of change under strain $s$. The hopping amplitudes and rates have been evaluated by DFT and are summarized in Table II in the main text. Motivated by previous work, we take $\lambda_{xz/xy} = t_{xz/xy}/10$ along the respective directions and we set the correlation enhanced SOC $\eta = 0.2eV$.

Analogously, we can parametrize the local gap matrices in the orbital basis as

$$\hat{\Delta}^{XZ} = d_0\,\tau_a \otimes \sigma_b\,(i\sigma_2). \tag{7}$$

The irreps associated with each $[a, b]_{XZ}$ are given in the first column of Table IV, with the following columns highlighting the spin and orbital character, and the respective order parameter in the original three-dimensional three-orbital model.

| 2-orbital model | | 3-orbital model | | | |
|---|---|---|---|---|---|
| Irrep in D$_{2h}$ | $(a,b)_{XZ}$ | Irrep in D$_{4h}$ | (a,b) | Physical process | |
| $A_g$ | $(0,0)_{XZ}$ | $A_{1g}$ | $(0,0)$ | intra-orbital hopping | |
| | $(3,0)_{XZ}$ | $A_{1g}/B_{1g}$ | $(8,0)/(7,0)$ | intra-orbital hopping | |
| | $(2,1)_{XZ}$ | $A_{1g}$ and $B_{1g}$ | $(6,1)$ | atomic SOC | |
| $B_{1g}$ | $(2,2)_{XZ}$ | $A_{2g}$ and $B_{2g}$ | $(6,2)$ | **k**-SOC | |
| $B_{2g}$ | $(2,3)_{XZ}$ | $E_g$ | $(6,3)$ | **k**-SOC | |
| $B_{3g}$ | $(1,0)_{XZ}$ | $E_g$ | $(3,0)$ | inter-orbital hopping | |

TABLE III. Symmetry-allowed terms in the normal state Hamiltonian along the $XZ$-plane.

| 2-orbital model | | 3-orbital model | | | |
|---|---|---|---|---|---|
| Irrep in $D_{2h}$ | $[a,b]_{XZ}$ | Irrep in $D_{4h}$ | $[a,b]$ | Orbital | Spin |
| $A_g$ | $[0,0]_{XZ}$ | $A_{1g}$ | $[0,0]$ | symmetric | singlet |
| | $[3,0]_{XZ}$ | $A_{1g}$ /$B_{1g}$ | $[8,0]/[7,0]$ | symmetric | singlet |
| | $[2,1]_{XZ}$ | $A_{1g}$ and $B_{1g}$ | $[6,1]$ | antisymmetric | triplet |
| $B_{1g}$ | $[2,2]_{XZ}$ | $A_{2g}$ and $B_{2g}$ | $[6,2]$ | antisymmetric | triplet |
| $B_{2g}$ | $[2,3]_{XZ}$ | $E_g$ | $[6,3]$ | antisymmetric | triplet |
| $B_{3g}$ | $[1,0]_{XZ}$ | $E_g$ | $[3,0]$ | symmetric | singlet |

TABLE IV. Local gap functions for the model projected along the $XZ$-plane.

## Model along the $YZ$ plane

An analogous construction gives us the model for the normal state Hamiltonian and order parameters along the $YZ$ plane. The symmetry allowed terms $(a, b)_{YZ}$ are the same, but are associated with different irreps and correspond to distinct physical processes, when compared to the model along the $XZ$ plane.

The explicit form of the parameters $\tilde{h}_{ab}^{YZ}(\mathbf{k})$ is:

$$\tilde{h}_{00}^{YZ}(\mathbf{k}) = [\xi_{yz}(\mathbf{k}) + \xi_{xy}(\mathbf{k})]/2 \tag{8}$$

$$\tilde{h}_{10}^{YZ}(\mathbf{k}) = \xi_{yz/xy}(\mathbf{k}) = 0 \qquad \text{at the } k_x = 0 \text{ plane***}$$

$$\tilde{h}_{21}^{YZ}(\mathbf{k}) = \eta_{yz/xy}^1(\mathbf{k}) = 0 \qquad \text{at the } k_x = 0 \text{ plane***}$$

$$\tilde{h}_{22}^{YZ}(\mathbf{k}) = \eta$$

$$\tilde{h}_{23}^{YZ}(\mathbf{k}) = \eta_{yz/xy}^3(\mathbf{k})$$

$$\tilde{h}_{30}^{YZ}(\mathbf{k}) = [\xi_{yz}(\mathbf{k}) - \xi_{xy}(\mathbf{k})]/2,$$

with

$$\xi_{yz}(\mathbf{k}) = 2t_{yz}(\hat{x})\cos(k_x a) + 2t_{yz}(\hat{y})\cos k_y a + 2t_{yz}(\hat{d}_p)\cos(k_x a + k_y a) + 2t_{yz}(\hat{\bar{d}}_p)\cos(k_x a - k_y a)$$

$$\xi_{xy}(\mathbf{k}) = 2t_{xy}(\hat{x})\cos(k_x a) + 2t_{xy}(\hat{y})\cos k_y a + 2t_{xy}(\hat{d}_p)\cos(k_x a + k_y a) + 2t_{xy}(\hat{\bar{d}}_p)\cos(k_x a - k_y a)$$

$$\xi_{yz/xy}(\mathbf{k}) = -4t_{yz/xy}(\hat{d}_d)\sin\left(\frac{k_x + k_y}{2}\right)\sin\left(\frac{k_z}{2}\right) - 4t_{yz/xy}(\hat{\bar{d}}_d)\sin\left(\frac{k_x - k_y}{2}\right)\sin\left(\frac{k_z}{2}\right)$$

$$\eta_{yz/xy}^1(\mathbf{k}) = -2\lambda_{yz/xy}(\hat{d}_p)\cos(k_x + k_y) + 2\lambda_{yz/xy}(\hat{\bar{d}}_p)\cos(k_x - k_y)$$

$$\eta_{yz/xy}^3(\mathbf{k}) = 4\lambda_{yz/xy}(\hat{d}_d)\sin\left(\frac{k_x + k_y}{2}\right)\sin\left(\frac{k_z}{2}\right) - 4\lambda_{yz/xy}(\hat{\bar{d}}_d)\sin\left(\frac{k_x - k_y}{2}\right)\sin\left(\frac{k_z}{2}\right).$$

Considerations discussed for the $XZ$ plane are also valid here.

The symmetry allowed local order parameters $[a, b]_{YZ}$ are also the same, but these are associated with different irreps and correspond to different components in the two-dimensional irrep in the original three-dimensional model. The results are summarized in Tables V and VI.

| 2-orbital model | | 3-orbital model | | | |
|---|---|---|---|---|---|
| Irrep in D$_{2h}$ | $(a,b)_{XZ}$ | Irrep in D$_{4h}$ | (a,b) | Physical process | |
| $A_g$ | $(0,0)_{YZ}$ | $A_{1g}$ | $(0,0)$ | intra-orbital hopping | |
| | $(3,0)_{YZ}$ | $A_{1g}$ /$B_{1g}$ | $(8,0)/(7,0)$ | intra-orbital hopping | |
| $B_{1g}$ | $(2,1)_{YZ}$ | $A_{2g}$ and $B_{2g}$ | $(5,1)$ | **k**-SOC | |
| $A_g$ | $(2,2)_{YZ}$ | $A_{1g}$ and $B_{1g}$ | $(5,2)$ | atomic SOC | |
| $B_{3g}$ | $(2,3)_{YZ}$ | $E_g$ | $(5,3)$ | **k**-SOC | |
| $B_{2g}$ | $(1,0)_{YZ}$ | $E_g$ | $(2,0)$ | inter-orbital hopping | |

TABLE V. Symmetry-allowed terms in the normal state Hamiltonian along the $YZ$-plane.

| 2-orbital model | | 3-orbital model | | | |
|---|---|---|---|---|---|
| Irrep in $D_{2h}$ | $[a,b]_{XZ}$ | Irrep in $D_{4h}$ | $[a,b]$ | Orbital | Spin |
| $A_g$ | $[0,0]_{YZ}$ | $A_{1g}$ | $[0,0]$ | symmetric | singlet |
| | $[3,0]_{YZ}$ | $A_{1g}$ /$B_{1g}$ | $[8,0]/[7,0]$ | symmetric | singlet |
| $B_{1g}$ | $[2,1]_{YZ}$ | $A_{2g}$ and $B_{2g}$ | $[5,1]$ | antisymmetric | triplet |
| $A_g$ | $[2,2]_{YZ}$ | $A_{1g}$ and $B_{1g}$ | $[5,2]$ | antisymmetric | triplet |
| $B_{3g}$ | $[2,3]_{YZ}$ | $E_g$ | $[5,3]$ | antisymmetric | triplet |
| $B_{2g}$ | $[1,0]_{YZ}$ | $E_g$ | $[2,0]$ | symmetric | singlet |

TABLE VI. Local gap functions for the model projected along the $YZ$-plane.

## SUPERCONDUCTING FITNESS ANALYSIS

Here we propose an analysis for the effects of strain based on the superconducting fitness measures [3, 4]. As the simplest form of these measures was introduced in the context of two-orbital models, here we focus on the reduced models along the $XZ$ and $YZ$ planes with $D_{2h}$ symmetry. Note that there are no two-dimensional irreps in $D_{2h}$, so at a first sight it seems that we have lost the discussion about the splitting of the degenerate superconducting transitions. Now the reduced model along the $XZ$ plane should be seen as degenerate with the reduced model along the $YZ$ plane, and their inequivalence under strain is a manifestation of the symmetry breaking.

Within the standard weak-coupling assumptions, the superconducting critical temperature for a two-orbital system can be written as [4]:

$$T_c(s) \approx \frac{4e^\gamma}{\pi} \frac{\omega_C}{2} e^{-1/2|v|\alpha(s)} e^{-\delta(s)/\alpha(s)},$$  (9)

where $\gamma$ is the Euler constant, $\omega_C$ is a characteristic energy cutoff, $|v|$ is the magnitude of the attractive interaction in the symmetry channel of interest, the last two assumed to be strain-independent. The superconducting fitness functions are:

$$\alpha(s) = \frac{1}{16} \sum_a N_a(0, s) \langle ||\hat{F}_A(\mathbf{k}_F, s)||^2 \rangle_{FS_a},$$  (10)

and

$$\delta(s) = \frac{\omega_C^2}{32} \sum_a N_a(0, s) \left\langle \frac{||\hat{F}_C(\mathbf{k}_F, s)||^2}{q(\mathbf{k}_F)^2} \right\rangle_{FS_a},$$  (11)

where $q(\mathbf{k}) = \epsilon_a(\mathbf{k}) - \epsilon_b(\mathbf{k})$ corresponds to the energy difference between the two bands. The sum over $a$ corresponds to the sum over the Fermi surfaces with DOS $N_a(0, s)$ at the Fermi level for strain $s$, and $\langle ... \rangle_{FS_a}$ corresponds to the average over the respective Fermi surface. $||\hat{M}||^2 = \text{Tr}[\hat{M}\hat{M}^\dagger]$ corresponds to the Frobenius norm of the matrix $\hat{M}$. The superconducting fitness matrices are defined as:

$$\hat{F}_A(\mathbf{k}, s)(i\sigma_2) = \tilde{H}_0(\mathbf{k}, s)\tilde{\Delta}(\mathbf{k}) + \tilde{\Delta}(\mathbf{k})\tilde{H}_0^*(-\mathbf{k}, s),$$  (12)

$$\hat{F}_C(\mathbf{k}, s)(i\sigma_2) = \tilde{H}_0(\mathbf{k}, s)\tilde{\Delta}(\mathbf{k}) - \tilde{\Delta}(\mathbf{k})\tilde{H}_0^*(-\mathbf{k}, s),$$  (13)

such that the effects of strain are introduced through the change in the Hamiltonian parameters. Here $\tilde{H}_0(\mathbf{k}, s) = [\hat{H}_0(\mathbf{k}, s) - \tilde{h}_{00}(\mathbf{k}, s)\sigma_0 \otimes \tau_0]/|\tilde{h}(\mathbf{k}, s)|$ corresponds to the normalized normal state Hamiltonian matrix without the term proportional to the identity, with $|\tilde{h}(\mathbf{k}, s)|^2 = \sum_{(a,b)\neq(0,0)} |\tilde{h}_{ab}(\mathbf{k}, s)|^2$. Here $\tilde{\Delta}(\mathbf{k}) = \hat{\Delta}(\mathbf{k})/d_0$ is the normalized gap matrix.

From Eq. 9 it becomes clear that a larger $\alpha(s)$ guarantees a larger critical temperature, while a larger $\delta(s)$ is associated with a smaller critical temperature. The evaluation of $\hat{F}_A(\mathbf{k}, s)$ and $\hat{F}_C(\mathbf{k}, s)$ for each local order parameter in the two-orbital models introduced above is summarized in Table VII.

| | $\hat{F}_A$ | | | | | $\hat{F}_C$ | | | | |
|---|---|---|---|---|---|---|---|---|---|---|
| | $(3,0)_p$ | $(2,1)_p$ | $(2,2)_p$ | $(2,3)_p$ | $(1,0)_p$ | $(3,0)_p$ | $(2,1)_p$ | $(2,2)_p$ | $(2,3)_p$ | $(1,0)_p$ |
| $[0,0]_p$ | 1 | 1 | 1 | 1 | 1 | 0 | 0 | 0 | 0 | 0 |
| $[3,0]_p$ | 1 | 0 | 0 | 0 | 0 | 0 | 1 | 1 | 1 | 1 |
| $[2,1]_p$ | 0 | 1 | 0 | 0 | 0 | 1 | 0 | 1 | 1 | 1 |
| $[2,2]_p$ | 0 | 0 | 1 | 0 | 0 | 1 | 1 | 0 | 1 | 1 |
| $[2,3]_p$ | 0 | 0 | 0 | 1 | 0 | 1 | 1 | 1 | 0 | 1 |
| $[1,0]_p$ | 0 | 0 | 0 | 0 | 1 | 1 | 1 | 1 | 1 | 0 |

TABLE VII. Superconducting fitness analysis for the effective two-orbital models along the planes $p = \{XZ, YZ\}$. The dependence of $\hat{F}_{A,C}(\mathbf{k}, s)$ on momenta $\mathbf{k}$ and strain $s$ is implicit through $\tilde{h}_{ab} = \tilde{h}_{ab}(\mathbf{k}, s)$. The table should be read as follows: Each line corresponds to an order parameter with matrix structure $[a, b]_p$ in plane $p$, and the numerical entries correspond to $||\hat{F}_{A,C}(\mathbf{k})||^2 = 4\sum_{cd}(\text{table entry})|\tilde{h}_{cd}(\mathbf{k})|^2/|\tilde{h}(\mathbf{k})|^2$, for each term $\tilde{h}_{cd}(\mathbf{k})$ in the normal-state Hamiltonian indicated in the second row as $(c, d)_p$. Note that $\hat{F}_A(\mathbf{k})$ and $\hat{F}_C(\mathbf{k})$ are complementary, with $||\hat{F}_A(\mathbf{k})||^2 + ||\hat{F}_C(\mathbf{k})||^2 = 4$. The irrep associated with each $[a, b]_p$ term and the associated order parameter in the original three-orbital model should be read from Tables IV and VI for each plane.

We can then use the relation $||\hat{F}_A(\mathbf{k}, s)||^2 + ||\hat{F}_C(\mathbf{k}, s)||^2 = 4$ and rewrite $\hat{F}_C(\mathbf{k}, s)$ as:

$$\delta(s) = \frac{\omega_C^2}{32} \sum_a N_a(0, s) \left\langle \frac{4 - ||\hat{F}_A(\mathbf{k}_F, s)||^2}{q(\mathbf{k}_F)^2} \right\rangle_{FS_a}. \tag{14}$$

Given the proximity of the two Fermi surfaces at the XZ and YZ planes, we assume that the superconducting fitness functions for each Fermi surface are going to be approximately the same. For simplicity, we take a single representative point in between the $\beta$ and $\gamma$ bands for our calculations. We also assume the difference between the bands $q(\mathbf{k})$ to be a constant over the Fermi surface and not to vary with strain. Under these assumptions, we can factor out the total density of states $N(0, s)$ and write:

$$\delta(s) \approx \frac{\omega_C^2}{32} N(0, s) \frac{4 - \langle ||\hat{F}_A(\mathbf{k}_F, s)||^2 \rangle_{FS}}{q(\mathbf{k}_F)^2}. \tag{15}$$

In terms of the unstrained value $\delta(s)$, we can write:

$$\delta(s) \approx \delta(0) \frac{N(0, s)}{N(0, 0)} \frac{\langle ||\hat{F}_C(\mathbf{k}_F, 0)||^2 \rangle_{FS} + \langle ||\hat{F}_A(\mathbf{k}_F, 0)||^2 \rangle_{FS} - \langle ||\hat{F}_A(\mathbf{k}_F, s)||^2 \rangle_{FS}}{\langle ||\hat{F}_C(\mathbf{k}_F, 0)||^2 \rangle_{FS}}. \tag{16}$$

In order to get more intuition from the quantitative discussion, we can expand the average of the fitness function $\hat{F}_A(\mathbf{k}_F, s)$ for small values of strain as:

$$\langle ||\hat{F}_A(\mathbf{k}_F, s)||^2 \rangle_{FS} \approx \langle ||\hat{F}_A(\mathbf{k}_F, 0)||^2 \rangle_{FS}(1 + F_1 s + F_2 s^2), \tag{17}$$

and the total density of states as:

$$N(0, s) \approx N(0, 0)(1 + N_1 s + N_2 s^2). \tag{18}$$

with coefficients $F_{1,2}$ and $N_{1,2}$ estimated from DFT calculations. Their values for strain along different directions are summarized in Table II in the main text.

With this parametrization, $\delta(s)$ can be written as:

$$\delta(s) \approx \delta(0) \frac{N(0, s)}{N(0, 0)} \left[ 1 - \frac{\langle ||\hat{F}_A(\mathbf{k}, 0)||^2 \rangle_{FS}}{\langle ||\hat{F}_C(\mathbf{k}, 0)||^2 \rangle_{FS}} (F_1 s + F_2 s^2) \right] \tag{19}$$

$$\approx \delta(0)(1 + N_1 s + N_2 s^2) \left[ 1 - A(F_1 s + F_2 s^2) \right].$$

Here we defined $A = \frac{\langle ||\hat{F}_A(\mathbf{k}, 0)||^2 \rangle_{FS_a}}{\langle ||\hat{F}_C(\mathbf{k}, 0)||^2 \rangle_{FS_a}}$. For the $[2, 3]_{XZ/YZ}$ order parameters, we estimate $A \approx 10^{-5}$, and for the $[1, 0]_{XZ/YZ}$ order parameters $A = 0$ (see quantitative discussion below). The small value of $A$ allows us to neglect the dependence of $\delta(s)$ on strain through the $F_{1,2}$ coefficients. Furthermore, within the assumption that $\frac{1}{2|v|} \gg \delta(0)$, we can take

$\delta(s) \approx \delta(0)$, such that the dependence of the critical temperature on strain is carried only by the fitness function $\alpha(s)$. Within these considerations, the closed form equation for $T_c$ then simplifies to:

$$T_c(s) \approx \frac{4e^\gamma}{\pi} \frac{\omega_C}{2} \exp\left[-1/2|v| - \delta(0)\right] / \alpha(s), \tag{20}$$

and we can finally write:

$$\begin{aligned} T_c(s) &\approx T_c(0) \exp\left[-1/2|v| - \delta(0)\right] (1/\alpha(s) - 1/\alpha(0)) \\ &\approx T_c(0) \exp\left[g\left(1 - \frac{\alpha(0)}{\alpha(s)}\right)\right], \end{aligned} \tag{21}$$

where $g = [1/(2|v| + \delta(0))]/\alpha(0)$, as defined in the main text.

## QUANTITATIVE ANALYSIS

### Analysis for the $\{[5,3],[6,3]\}$ order parameter

We now discuss how the superconducting fitness parameters evolve as a function of strain for the order parameter components $\{[5,3],[6,3]\}$ based on the behaviour of the dominant hopping amplitudes estimated from DFT calculations (results summarized in Table VIII). This analysis gives rise to the coefficients $F_1$ and $F_2$ summarized in Table II in the main text. Here it is important to remember the mapping to the projected two-orbital models as $[5,3] \to [2,3]_{YZ}$ and $[6,3] \to [2,3]_{XZ}$.

From Table I in the main text, for the order parameter $[2,3]_p$, the normal state Hamiltonian term $(2,3)_p$ contributes to a finite $\hat{F}_A(\mathbf{k})$, and the remaining contribute to a finite $\hat{F}_C(\mathbf{k})$. Writing explicitly:

$$\begin{aligned} ||\hat{F}_A(\mathbf{k},s)||^2 &= 4|\tilde{h}_{23}(\mathbf{k},s)|^2/|\tilde{h}(\mathbf{k},s)|^2, \\ ||\hat{F}_C(\mathbf{k},s)||^2 &= 4[|\tilde{h}_{30}(\mathbf{k},s)|^2 + |\tilde{h}_{21}(\mathbf{k},s)|^2 + |\tilde{h}_{22}(\mathbf{k},s)|^2 + |\tilde{h}_{10}(\mathbf{k},s)|^2]/|\tilde{h}(\mathbf{k},s)|^2. \end{aligned} \tag{22}$$

Focusing first on the $XZ$ plane, the term $(3,0)_{XZ}$ is associated with the imbalance in the intra-orbital nearest-neighbour hopping amplitudes for the $xz$ and $xy$ orbitals, and is proportional to $t_{xz}(\hat{x}) - t_{xy}(\hat{x}) \approx -0.06eV$; $(2,1)_{XZ}$ is associated with the correlation-enhanced effective atomic SOC, $\eta \approx 0.2eV$ [5]; $(2,3)_{XZ}$ is associated with momentum-dependent SOC in the $B_{2g}$ irrep with magnitude much smaller than the atomic SOC. As an estimate, here we take its value to be equal to one tenth the hopping amplitude with

the same inter-orbital structure [2], $t_{xy/xz}^{SOC}(\hat{d}_d) \approx t_{xy/xz}(\hat{d}_d)/10 \approx 0.68meV$. By symmetry, $(2,2)_{XZ}$ and $(1,0)_{XZ}$ are zero along the $XZ$ plane. We therefore focus on the evolution of the terms $(3,0)_{XZ}$ and $(2,3)_{XZ}$ with strain. The evolution of the hopping amplitudes as a function of strain is determined by DFT calculations, performed without SOC. The correlation-enhanced effective atomic SOC is taken by its agreement with the experimental Fermi surfaces [5] and was verified not to change significantly with strain (this result will be published elsewhere).

Analogously, for the $YZ$ plane, the term $(3,0)_{YZ}$ is associated with the imbalance in the intra-orbital nearest-neighbour hopping amplitudes for the $yz$ and $xy$ orbitals, and is proportional to $t_{yz}(\hat{y}) - t_{xy}(\hat{y}) \approx -0.06eV$; $(2,2)_{YZ}$ is associated with the effective atomic SOC, $\eta \approx 0.2eV$ [5], which is not expected to change under strain; $(2,3)_{YZ}$ is associated with momentum-dependent SOC in the $B_{3g}$ irrep with magnitude much smaller than the effective atomic SOC. As an estimate, here we take its value to be equal to one tenth the hopping amplitude with the same inter-orbital structure [2], $t_{xy/yz}^{SOC}(\hat{d}_d) \approx t_{xy/yz}(\hat{d}_d)/10 \approx 0.68meV$. By symmetry, $(2,1)_{YZ}$ and $(1,0)_{YZ}$ are zero along the $YZ$ plane. We therefore focus on the evolution of the terms $(3,0)_{YZ}$ and $(2,3)_{YZ}$ with strain.

*Strain along the $\langle 001 \rangle$ direction*

For compressive strain along the $\langle 001 \rangle$ direction, there is no reduction of the symmetry group and, therefore, no splitting of the superconducting transition temperatures. Considering the model projected along the $XZ$-plane, $(3,0)_{XZ} \sim t_{xz}(\hat{x}) - t_{xy}(\hat{x}) \approx -0.065eV$ is enhanced in magnitude, leading to an increase of $\hat{F}_C(\mathbf{k})$, and $(2,3)_{YZ}$ is reduced, leading to a reduction in $\hat{F}_A(\mathbf{k})$, such that the critical temperature is expected to be reduced for the $[6,3]$ component, if the coefficient $F_1$ dominates over the coefficient $N_1$ for small strain. The analysis along the $YZ$-plane leads to the same conclusion for the $[5,3]$ component. The result is displayed as the red curve in Fig. 1 (a) in the main text. Note that the value of $g$, chosen to match the critical temperature for $s = -0.5\%$ along the $\langle 100 \rangle$ direction, provides good qualitative agreement for strain along the $\langle 001 \rangle$ direction.

## Analysis for the $\{[3,0],-[2,0]\}$ order parameter

We now discuss how the superconducting fitness parameters evolve as a function of strain for the second superconducting order parameter associated with a two-dimensional irreducible representation in the original model, $\{[3,0],-[2,0]\}$. Here is important to remember the mapping to the projected two-orbital models as $[3,0] \to [1,0]_{XZ}$ and $[2,0] \to [1,0]_{YZ}$.

From Table I in the main text, for the order parameter $[1,0]_p$, the normal state Hamiltonian term $(1,0)_p$ contributes to a finite $\hat{F}_A(\mathbf{k})$, and the remaining contribute to a finite $\hat{F}_C(\mathbf{k})$. Writing explicitly:

$$||\hat{F}_A(\mathbf{k},s)||^2 = 4|\tilde{h}_{10}(\mathbf{k},s)|^2/|\tilde{h}(\mathbf{k},s)|^2, \tag{23}$$
$$||\hat{F}_C(\mathbf{k},s)||^2 = 4[|\tilde{h}_{30}(\mathbf{k},s)|^2 + |\tilde{h}_{21}(\mathbf{k},s)|^2 + |\tilde{h}_{22}(\mathbf{k},s)|^2 + |\tilde{h}_{23}(\mathbf{k},s)|^2]/|\tilde{h}(\mathbf{k},s)|^2.$$

Note that, by symmetry, $\tilde{h}_{10}(\mathbf{k}) = 0$ along the $XZ$ and $YZ$ planes, such that $||\hat{F}_{A,C}(\mathbf{k},s)||^2$ does not change under strain, and the evolution of the critical temperature is then purely controlled by the evolution of the DOS. The evolution of the critical temperatures under strain along the different directions for $\{[3,0],-[2,0]\}$ order parameter is displayed in Fig. 1 (b) in the main text. Note that now compressive strain along the $\langle 001 \rangle$ direction leads to an enhancement of the critical temperature.

## Details of the DFT calculation

For the $<100>$ and $<001>$ strain directions, the DFT calculations were performed with Wien2k [6] using RKmax=8, in the generalized gradient approximation (GGA) for the exchange-correlation functional as parametrized by Perdew, Burke, and Ernzerhof (PBE) [7]. The structure was taken as the experimental lattice parameters $a = b = 3.8613 \mathring{A}$ , $c = 12.7218 \mathring{A}$ from Ref.[8] measured at 8 K. The internal coordinates (apical O and Sr z-position) of the unstrained structure were relaxed within DFT resulting in $z_O = 0.16344$ and $z_{Sr} = 0.35230$ (compared to the experimental ones of: $0.1634(4)$ and $0.3529(4)$). For simplicity we do not optimize the internal coordinates under strain. We confirmed with calculations at higher strains that the internal coordinates are not affected much by the applied strain. We apply strain in the $<100>$ direction (i.e. the $<001>$ direction) and scale the $<010>$ and $<001>$ directions according to the experimental Poisson ratios of

$\epsilon_{yy}$ / $\epsilon_{xx}$ = 0.508 and $\epsilon_{zz}$ / $\epsilon_{xx}$ = 0.163 [9]. The calculations are performed on a shifted $27 \times 27 \times 27$ $k$-grid for the primitive 7-atom unit cell.

For strain along < 110 > direction, the DFT calculations were performed using the Quantum ESPRESSO package [10] using GGA-PBE. Scalar-relativistic ultrasoft pseudopotential from the GBRV library [11] were used, with the 4s and 4p semicore states of Sr and Ru included in the valence. The plane wave kinetic energy cutoff is set to 60 Ry, and the Brillouin-zone integration is carried out using a Methfessel-Paxton smearing parameter of 0.01 Ry. To implement the uniaxial strain along < 110 >, the conventional 14-atom unit-cell is enlarged with $\sqrt{2} \times \sqrt{2}$ in-plane lattice vectors, resulting in a 28-atom supercell with one in-plane axes aligned along the < 110 > direction. For the structural optimization, compressive strain is imposed on this lattice parameter, while the two perpendicular lattice parameters are relaxed until the corresponding components of the stress tensor are smaller than 0.1 kb. As before, the internal coordinates are kept fixed during the relaxation, calculated on a $\Gamma$-centered $7 \times 7 \times 3$ $k$-point grid for the supercell. The charge density is recalculated for the primitive 7-atom unit cell on a $\Gamma$-centered $27 \times 27 \times 27$ $k$-point grid.

To describe the low-energy bands crossing the Fermi energy we construct three maximally localized Wannier functions [12, 13] of $t_{2g}$ symmetry using wien2wannier [14] and Wannier90 [15], a 10 x 10 x 10 k-grid and a frozen energy window ranging from -1.77 to 3 eV for all strains. From the real-space Wannier model the relevant hopping amplitudes are extracted and the strain-dependent rates calculated from a first-order fit with strain up to 0.8% (summarized in Table VIII). To estimate the coefficients of the evolution of the density of states at the Fermi energy in the $k_x k_z$ and $k_y k_z$ planes, the eigenstates are re-evaluated on a planar mesh of 19600 $k$-points using the TRIQS library [16]. To remove a remaining influence of the limited resolution, the density of states at the Fermi level is determined from a third-order fit in an energy range of up to $\pm 0.05$ eV. Based on the resulting values N(0,s) as function of strain, the coefficients $N_1$ and $N_2$ are obtained from a second-order fit for the $k_x k_z$ and $k_y k_z$ plane, respectively (summarized in Table VIII).

| | ⟨100⟩ | | ⟨001⟩ | | ⟨110⟩ | |
|---|---|---|---|---|---|---|
| Hopping | Amplitude | Rate | Amplitude | Rate | Amplitude | Rate |
| $t_{xy}(\hat{x})$ | 0.37344 | -0.02243 | 0.37354 | 0.00060 | 0.36288 | -0.00332 |
| $t_{xy}(\hat{y})$ | 0.37364 | 0.01439 | 0.37354 | 0.00060 | 0.36288 | -0.00332 |
| $t_{xy}(\hat{d}_p)$ | 0.12208 | -0.00138 | 0.12207 | 0.00131 | 0.11981 | -0.00332 |
| $t_{xy}(\hat{\bar{d}}_p)$ | 0.12208 | -0.00138 | 0.12207 | 0.00131 | 0.11982 | 0.00090 |
| $t_{xz}(\hat{x})$ | 0.30895 | -0.02304 | 0.30837 | 0.01372 | 0.30777 | -0.00860 |
| $t_{xz}(\hat{y})$ | 0.04300 | 0.00219 | 0.04298 | -0.00201 | 0.04015 | 0.00004 |
| $t_{xz}(\hat{d}_p)$ | 0.01604 | 0.00092 | 0.1610 | -0.00067 | 0.01499 | 0.00016 |
| $t_{xz}(\hat{\bar{d}}_p)$ | 0.01604 | 0.00092 | 0.01610 | -0.00067 | 0.01499 | 0.00008 |
| $t_{yz}(\hat{x})$ | 0.04300 | -0.00220 | 0.04298 | -0.00201 | 0.04015 | 0.00004 |
| $t_{yz}(\hat{y})$ | 0.30909 | 0.00373 | 0.30837 | 0.01372 | 0.30777 | -0.00860 |
| $t_{yz}(\hat{d}_p)$ | 0.01604 | -0.00059 | 0.01610 | -0.00067 | 0.01499 | 0.00016 |
| $t_{yz}(\hat{\bar{d}}_p)$ | 0.01604 | -0.00059 | 0.01610 | -0.00067 | 0.01499 | 0.00008 |
| $t_{xy/xz}(\hat{d}_d)$ | 0.00678 | -0.00016 | 0.00678 | -0.00033 | 0.00633 | 0.00021 |
| $t_{xy/xz}(\hat{\bar{d}}_d)$ | 0.00678 | -0.00016 | 0.00678 | -0.00033 | 0.00633 | -0.00016 |
| $t_{xy/yz}(\hat{d}_d)$ | 0.00677 | 0.00023 | 0.00678 | -0.00033 | 0.00633 | 0.00021 |
| $t_{xy/yz}(\hat{\bar{d}}_d)$ | 0.00677 | 0.00023 | 0.00678 | -0.00033 | 0.00633 | -0.00016 |

TABLE VIII. Hopping amplitudes (in eV) and rate of change (in eV/%) under strain along ⟨100⟩, ⟨001⟩, and ⟨110⟩ directions. Here $t_\gamma(\hat{n})$ corresponds to the intra-orbital hopping amplitude for orbital $\gamma$ and $t_{\gamma/\gamma'}(\hat{n})$ corresponds to inter-orbital hopping between orbitals $\gamma$ and $\gamma'$ in direction $\hat{n}$. The rates of change of the hopping amplitudes on this table are for compressive strain corresponding to $s > 0$. Estimates based on DFT calculations for compressive strains up to -1.6% along the ⟨100⟩ direction and up to -3.2% along the ⟨001⟩ direction. Note that compressive strain corresponds to negative values.

[1] A. Ramires and M. Sigrist, Phys. Rev. B **100**, 104501 (2019).

[2] H. G. Suh, H. Menke, P. M. R. Brydon, C. Timm, A. Ramires, and D. F. Agterberg, Phys. Rev. Research **2**, 032023 (2020).

[3] A. Ramires and M. Sigrist, Phys. Rev. B **94**, 104501 (2016).

[4] A. Ramires, D. F. Agterberg, and M. Sigrist, Phys. Rev. B **98**, 024501 (2018).

[5] A. Tamai, M. Zingl, E. Rozbicki, E. Cappelli, S. Riccò, A. de la Torre, S. McKeown Walker, F. Y. Bruno, P. D. C. King, W. Meevasana, M. Shi, M. Radović, N. C. Plumb, A. S. Gibbs, A. P. Mackenzie, C. Berthod, H. U. R. Strand, M. Kim, A. Georges, and F. Baumberger, Phys. Rev. X **9**, 021048 (2019).

[6] P. Blaha, K. Schwarz, G. K. HMadsen, D. Kvasnicka, J. Luitz, R. Laskowsk, F. Tran, L. Marks, and L. Marks, *WIEN2k: An Augmented Plane Wave Plus Local Orbitals Program for Calculating Crystal Properties.*

[7] J. P. Perdew, K. Burke, and M. Ernzerhof, Phys. Rev. Lett. **77**, 3865 (1996).

[8] T. Vogt and D. J. Buttrey, Phys. Rev. B **52**, R9843 (1995).

[9] M. E. Barber, F. Lechermann, S. V. Streltsov, S. L. Skornyakov, S. Ghosh, B. J. Ramshaw, N. Kikugawa, D. A. Sokolov, A. P. Mackenzie, C. W. Hicks, and I. I. Mazin, Phys. Rev. B **100**, 245139 (2019).

[10] P. Giannozzi, S. Baroni, N. Bonini, M. Calandra, R. Car, C. Cavazzoni, D. Ceresoli, G. L. Chiarotti, M. Cococcioni, I. Dabo, A. D. Corso, S. de Gironcoli, S. Fabris, G. Fratesi, R. Gebauer, U. Gerstmann, C. Gougoussis, A. Kokalj, M. Lazzeri, L. Martin-Samos, N. Marzari, F. Mauri, R. Mazzarello, S. Paolini, A. Pasquarello, L. Paulatto, C. Sbraccia, S. Scandolo, G. Sclauzero, A. P. Seitsonen, A. Smogunov, P. Umari, and R. M. Wentzcovitch, Journal of Physics: Condensed Matter **21**, 395502 (2009).

[11] K. F. Garrity, J. W. Bennett, K. M. Rabe, and D. Vanderbilt, Computational Materials Science **81**, 446 (2014).

[12] N. Marzari and D. Vanderbilt, Phys. Rev. B **56**, 12847 (1997).

[13] I. Souza, N. Marzari, and D. Vanderbilt, Phys. Rev. B **65**, 035109 (2001).

[14] J. Kuneš, R. Arita, P. Wissgott, A. Toschi, H. Ikeda, and K. Held, Computer Physics Communications **181**, 1888 (2010).

[15] G. Pizzi, V. Vitale, R. Arita, S. Blügel, F. Freimuth, G. Géranton, M. Gibertini, D. Gresch, C. Johnson, T. Koretsune, J. Ibañez-Azpiroz, H. Lee, J.-M. Lihm, D. Marchand, A. Marrazzo, Y. Mokrousov, J. I. Mustafa, Y. Nohara, Y. Nomura, L. Paulatto, S. Poncé, T. Ponweiser, J. Qiao, F. Thöle, S. S. Tsirkin, M. Wierzbowska, N. Marzari, D. Vanderbilt, I. Souza, A. A. Mostofi,  and J. R. Yates, Journal of Physics: Condensed Matter **32**, 165902 (2020).

[16] O. Parcollet, M. Ferrero, T. Ayral, H. Hafermann, I. Krivenko, L. Messio,  and P. Seth, Computer Physics Communications **196**, 398  (2015).



\end{document}